\documentclass[fleqn,usenatbib,useAMS]{mnras}
\usepackage{hyperref, orcidlink, multirow}
\usepackage{times, amsmath, amssymb, amsfonts, latexsym}
\usepackage{graphicx, color, xcolor, enumitem, ulem}
\usepackage{array, mathtools, esint, multicol, caption}
\usepackage{soul, bm, pdflscape}
\DeclareCaptionLabelFormat{noname}{#2}

\makeatletter
\@for\next:={int,iint,iiint,iiiint,dotsint,oint,oiint,sqint,sqiint,
	ointctrclockwise,ointclockwise,varointclockwise,varointctrclockwise,
	fint,varoiint,landupint,landdownint}\do{%
	\expandafter\edef\csname\next\endcsname{%
		\noexpand\DOTSI
		\expandafter\noexpand\csname\next op\endcsname
		\noexpand\ilimits@
	}%
}
\makeatother

\graphicspath{{./}{figures/}}

\title[Baryonic effects on the matter distribution]{
How do baryonic effects on the cosmic matter distribution vary with scale and local density environment?
}
\author[Y. Wang et al. ]{Yun Wang\orcidlink{0000-0003-4064-417X}$^1$\thanks{E-mail: yunw@jlu.edu.cn} and Ping He\orcidlink{0000-0001-7767-6154}$^{1,2}$\thanks{E-mail: hep@jlu.edu.cn} \\
	$^{1}$College of Physics, Jilin University, Changchun 130012, China. \\
	$^{2}$Center for High Energy Physics, Peking University, Beijing 100871, China.}

\date{Accepted 2024 January 18. Received 2024 January 18; in original form 2023 October 31}
\begin{document}
\maketitle

\begin{abstract}
In this study, we investigate how the baryonic effects vary with scale and local density environment mainly by utilizing a novel statistic, the environment-dependent wavelet power spectrum (env-WPS). With four state-of-the-art cosmological simulation suites, EAGLE, SIMBA, Illustris, and IllustrisTNG, we compare the env-WPS of the total matter density field between the hydrodynamic and dark matter–only (DMO) runs at $z=0$. We find that the clustering is most strongly suppressed in the emptiest environment of $\rho_\mathrm{m}/\bar\rho_\mathrm{m}<0.1$ with maximum amplitudes $\sim67-89$ per cent on scales $\sim1.86-10.96\ h\mathrm{Mpc}^{-1}$, and less suppressed in higher density environments on small scales (except Illustris). In the environments of $\rho_\mathrm{m}/\bar\rho_\mathrm{m}\geqslant0.316$ ($\geqslant10$ in EAGLE), the feedbacks also lead to enhancement features at intermediate and large scales, which is most pronounced in the densest environment of $\rho_\mathrm{m}/\bar\rho_\mathrm{m}\geqslant100$ and reaches a maximum $\sim 7-15$ per cent on scales $\sim0.87-2.62\ h\mathrm{Mpc}^{-1}$ (except Illustris). The baryon fraction of the local environment decreases with increasing density, denoting the feedback strength, and potentially explaining some differences between simulations. We also measure the volume and mass fractions of local environments, which are affected by $\gtrsim 1$ per cent due to baryon physics. In conclusion, our results show that the baryonic processes can strongly modify the overall cosmic structure on the scales of $k>0.1\ h\mathrm{Mpc}^{-1}$, which encourages further research in this direction.
\end{abstract}

\begin{keywords}
	cosmology: theory -- large-scale structure of Universe -- methods: numerical
\end{keywords}

\section{Introduction}
\label{sec:intro}

The standard $\Lambda$CDM cosmological paradigm depicts the geometry and matter composition of the Universe and predicts how structures like galaxies, galaxy clusters, and the cosmic web formed from tiny density fluctuations, which are confirmed by extensive observations 
\citep[e.g., see][for a review]{Frenk2012}. 
However, the collisionless cold dark matter-only (DMO) simulations indicate that the standard $\Lambda$CDM model faces some challenges at galactic and sub-galactic scales, such as the core-cusp problem, too big to fail problem, missing satellites problem, and diversity problem \citep[e.g.][]{Klypin1999, Blok2010, Boylan-Kolchin2011, Primack2012, Oman2015, Weinberg2015, Bullock2017}. The possible ways to solve these small-scale problems can be summarized into three categories: (i) replace DMO $N$-body simulations with hydrodynamic simulations incorporating baryons and baryonic physics \citep[e.g.][]{Sales2022}, (ii) propose alternatives to cold dark matter \citep[e.g.][]{Polisensky2011}, or (iii) modify the laws of gravity \citep[e.g.][]{Sanders2002}, in which the first solution holds the most promise, and alleviates the tension between theory and observation considerably \citep[e.g.][]{Chan2015, Simpson2018, Engler2021}. 

In order to faithfully mimic galaxy formation and evolution, the cosmological hydrodynamic simulations solve the Euler equations to simulate baryons, and adopt fine-tuned sub-grid (or sub-resolution) models to implement baryonic processes including, e.g., radiative cooling, star formation, and feedbacks from stellar winds, supernovae and active galactic nucleus (AGN) \citep[see][for review]{Vogelsberger2020,Crain2023}. Radiative cooling and star formation cause baryons to flow into the dark matter halo and collapse within the central region, whereas the feedbacks expel mass and energy of baryons from the halo. Moreover, the turbulent heating of the intergalactic medium may also be crucial in preventing the gravitational collapse of baryons \citep[e.g.][]{Zhu2010, Zhuravleva2014, Schmidt2016, Schmidt2017, Yang2020, Yang2022}. The inclusion of baryons and baryonic physics is anticipated to affect the spatial distribution of matter, partly due to the redistribution of baryons by baryonic processes, and partly due to the redistribution of dark matter by the gravitational coupling of baryons and dark matter, which is dubbed “back-reaction” \citep{vanDaalen2011}. Such baryonic effects on the cosmic matter distribution are still less well understood, which will lead to significant systematic uncertainties in the next-generation large-scale surveys (e.g. DESI\footnote{\url{https://desi.lbl.gov/}}, EUCLID\footnote{\url{http://sci.esa.int/euclid/}}, LSST\footnote{\url{https://www.lsst.org/}}, and WFIRST\footnote{\url{https://wfirst.gsfc.nasa.gov/}}), and therefore need to be carefully studied to extract unbiased cosmological information from survey data.

Numerous studies have made concerted efforts to investigate baryonic effects at low redshifts. These studies employed a variety of simulations with different box sizes, resolutions, cosmological parameters, and sub-grid models. Despite this diversity, the majority of these studies yield qualitatively consistent findings.  For instance, studies focusing on the halo shapes demonstrate that the presence of baryons leads to rounder halos \citep[e.g.][]{Schaller2015, Butsky2016, Henson2017, Chua2019,Chua2022, Cataldi2021, Cataldi2023}, and results on the halo mass function suggest that feedbacks can reduce the halo mass \citep[e.g.][]{Cui2014,Castro2021,BeltzMohrmann2021,Sorini2022}. In addition, baryonic processes greatly impact the matter power spectrum. Particularly, at redshift zero, the AGN feedback suppresses the power on scales of $0.1 h \mathrm{Mpc}^{-1}\lesssim k\lesssim20 h \mathrm{Mpc}^{-1}$ with the suppression reaching a maximum of $>10$ per cent at $k\sim10 h \mathrm{Mpc}^{-1}$, while on scales of $k\gtrsim20 h\mathrm{Mpc}^{-1}$, the suppression begins to be less effective due to gas cooling \citep[e.g.][]{vanDaalen2011, vanDaalen2020, Hellwing2016, Chisari2018, Springel2018, Debackere2020, Arico2020}. However, we notice that turbulent heating can also suppress the matter power spectrum and alleviate the over-cooling problem \citep{Voit2005}, which is supported by many studies \citep[e.g.][]{Zhu2010, Zhuravleva2014, Schmidt2016, Schmidt2017, Yang2020,Yang2022}, while not being sufficiently emphasized by the leading simulations. Furthermore, considering that power spectrum cannot carry the full information of the matter density field, some works investigated baryonic effects on the higher order statistic, i.e. the bispectrum, and found that the equilateral bispectrum is enhanced around $k\sim2-3 h\mathrm{Mpc}^{-1}$ due to baryons backreacting on the dark matter within massive haloes \citep{Foreman2020, Takahashi2020, Arico2021}. It should be noted that most works mentioned above predict or explain the baryonic effects based on the strong assumption that baryons only alter the matter distribution within halos. By measuring the probability distribution function of the cosmic density field, however, the recent work of \citet{Sunseri2023} found that feedback processes can also have pronounced effects on voids, walls, and filaments, which highlights the importance of modeling baryonic effects in the whole cosmic structure. 

Motivated by previous studies, we intend to delve deeper into how the impact of baryonic physics on the matter distribution varies with scale and local environment. To this end, we adopt the environment-dependent wavelet power spectrum (env-WPS) as our fiducial statistic, which was originally proposed by \citet{Wang2022b}. As a bivariate function of scale and local density environment, the env-WPS is built upon the continuous wavelet transform (CWT) and tells how the clustering strength of matter is distributed over different environments and scales. In \citet{Wang2022b}, we demonstrated that the env-WPS is capable of discriminating between the non-Gaussian density field and the homogeneous Gaussian random field, while the traditional Fourier power spectrum (FPS) is helpless in this regard. Accordingly, the env-WPS is a more advanced and powerful tool to characterize the large-scale structure of the Universe.

To ensure the reliability of our results, we utilize four modern cosmological-scale simulation projects, EAGLE\footnote{\url{https://icc.dur.ac.uk/Eagle/index.php}}, SIMBA\footnote{\url{http://simba.roe.ac.uk/}}, Illustris\footnote{\url{https://www.illustris-project.org/}} and IllustrisTNG\footnote{\url{https://www.tng-project.org/}}. In this work, we focus on analyzing the total matter density field at $z=0$, which is highly nonlinear and non-Gaussian. By measuring and comparing the env-WPS of the total matter between the hydrodynamic and the corresponding DMO runs, we will quantify the scale- and environment-dependence of the effects of baryonic physics on the matter spatial distribution. Therefore, we hope that the present study will shed new light on the modeling baryonic effects.

The remainder of this manuscript is organized as follows. In Section \ref{sec:simulations}, we briefly describe the simulations used here. In Section\ref{sec:methods}, we introduce the env-WPS and its computational procedure. We present and discuss the results in Section \ref{sec:results}, followed by the summary and outlook in Section \ref{sec:summary}.

\newcommand\Tstrut{\rule{0pt}{2.6ex}}         
\newcommand\Bstrut{\rule[-0.9ex]{0pt}{0pt}}   
\begin{table*}
	\centering
	\caption{Simulation data employed in this study. From left to right, we present the simulation name, code used to conduct simulations, choice of cosmological parameters, redshift range, box size, run name, number of dark matter particles ($N_\mathrm{dm}$), initial number of gas cells ($N_\mathrm{gas}$ ), dark matter mass resolution ($m_\mathrm{dm}$), and baryon mass resolution ($m_\mathrm{gas}$).}
	\label{tab:simulations}
	\noindent
	\resizebox{\textwidth}{!}{
		\begin{tabular}{@{}llllcllccc}
			\hline
			Simulation & Code & Cosmology & $z$ Range & Box size & Run & $N_\mathrm{dm}$ &  $N_\mathrm{gas}$ & $m_\mathrm{dm}$  & $m_\mathrm{gas}$ \Tstrut\\
			& & & & ($h^{-1}\text{Mpc}$) & & & & ($M_\odot$) & ($M_\odot$) \Bstrut\\ \\[-1.2em]
			\hline
			\multirow{2}{*}{Illustris} & 
			\multirow{2}{*}{AREPO}    & \multirow{2}{2.6cm}{WMAP-9 \citep{Hinshaw2013}} & \multirow{2}{*}{$127 - 0$} & \multirow{2}{*}{$75$} & Illustris-1 & $1820^3$ & $1820^3$ & $6.3\times10^6$ & $1.3\times10^6$  \Tstrut\\ \\[-0.5em]
			 &   & &  &  & Illustris-1-Dark & $1820^3$ & -- & $7.5\times10^6$ & -- \Bstrut\\ 
			\hline
			\multirow{4}{*}{IllustrisTNG} & \multirow{4}{*}{AREPO}  & \multirow{4}{2.6cm}{Planck2015 \citep{Planck2015}} & \multirow{4}{*}{$127 - 0$} &  \multirow{2}{*}{$75$} & TNG100-1 & $1820^3$ & $1820^3$ & $7.5\times10^6$ & $1.4\times10^6$  \Tstrut\\ \\[-0.5em]
			& & & &  & TNG100-1-Dark & $1820^3$ & -- & $8.9\times10^6$ & --  \\ \\[-0.5em]
			& & & & \multirow{2}{*}{$205$} & TNG300-1 & $2500^3$ & $2500^3$ & $5.9\times10^7$ & $1.1\times10^7$  \\ \\[-0.5em]
			& & & &  & TNG300-1-Dark & $2500^3$ & -- & $7.0\times10^8$ & -- \Bstrut\\ 
			\hline
			\multirow{2}{*}{EAGLE} & \multirow{2}{1.6cm}{modified GADGET-3 SPH} & \multirow{2}{2.6cm}{Planck2013 \citep{Planck2013}} & \multirow{2}{*}{$127 - 0$} & \multirow{2}{*}{$67.77$} & RefL0100N1504 & $1504^3$ & $1504^3$ & $9.7\times10^6$ & $1.8\times10^6$  \Tstrut\\ \\[-0.5em]
			& & & & & DMONLYL0100N1504 & $1504^3$ & -- & $1.2\times10^7$ & -- \\ \\[-0.8em]
			\hline
			\multirow{2}{*}{SIMBA} & \multirow{2}{*}{GIZMO}	& \multirow{2}{2.6cm}{Planck2015 \citep{Planck2015}} & \multirow{2}{*}{$249 - 0$} & \multirow{2}{*}{$100$} & m100n1024 & $1024^3$ & $1024^3$ & $9.6\times10^7$ & $1.8\times10^7$  \Tstrut\\ \\[-0.5em]
			& & & & & m100n1024-DMO & $1024^3$ & -- & $1.1\times10^8$ & --  \\ \\[-0.8em]
			\hline
		\end{tabular}
	}
\end{table*}

\section{Simulations}
\label{sec:simulations}

For our purpose, we rely on four state-of-the-art cosmological hydrodynamic simulation suites: EAGLE \citep{Crain2015, Schaye2015, EAGLE2017}, SIMBA \citep{Dave2019}, Illustris \citep{Vogelsberger2013, Vogelsberger2014, Genel2014, Nelson2015}, and its successor IllustrisTNG \citep{Marinacci2018, Naiman2018, Springel2018, Nelson2018, Nelson2019, Pillepich2018a}, whose redshift snapshots are publicly available.

These simulations track the evolution of billions of dark matter particles and baryonic elements over large cosmological volumes with $L_\mathrm{box}\gtrsim 100\ \mathrm{Mpc}$ from the high-redshift ($z>100$) to the present ($z=0$). To do so, the Illustris and IllustrisTNG are executed with the Voronoi moving-mesh code AREPO \citep{Springel2010}, while the EAGLE and SIMBA are implemented using a modified version of the smoothed particle hydrodynamics code GADGET-3 \citep{Springel2005} and the
meshless finite-mass hydrodynamics code GIZMO \citep{Hopkins2015}, respectively. All the considered simulations assume a concordance $\Lambda$CDM cosmology with parameters fixed by WMAP or Planck satellites. They all take into account the radiative cooling of gas, the formation of stars from the cold gas, stellar evolution, black hole formation and accretion, stellar and AGN feedbacks, but with different implementations. For details of these simulations, the reader is referred to the literature \citep{Crain2015, Schaye2015, Dave2019, Vogelsberger2013, Torrey2014, Pillepich2018b, Weinberger2018}. In particular, AGN feedback plays a vital role in the simulations, which can regulate star formation, thereby alleviating the overcooling problem substantially \citep{McCarthy2010,McCarthy2011}. The Illustris simulation employs an over-violent radio-mode AGN feedback model, resulting in the gas fraction in group-sized haloes being lower than observed \citep{Genel2014}. Comparatively, IllustrisTNG, EAGLE, and SIMBA are more realistic simulations, yielding better consistency with the available observations of galaxy formation and
evolution.

In this work, we use the highest resolution runs of simulations at $z=0$, which are listed in Table \ref{tab:simulations}. Throughout this paper, we refer to Illustris-1, TNG100-1, TNG300-1, RefL0100N1504, and m100n1024 as Illustris, TNG100, TNG300, EAGLE and SIMBA, correspondingly. Comparing results from them enables us to scrutinize the effects of baryonic physics on the matter distribution in a less biased manner.

\section{Methods}
\label{sec:methods}
\subsection{The env-WPS of the density field}

The total matter distribution of the Universe can be represented by the density contrast, defined as
\begin{equation}
	\label{eq:density_constrast}
	\delta_\mathrm{m}(\mathbf{x})\equiv\rho_\mathrm{m}(\mathbf{x})/\bar\rho_\mathrm{m}-1,
\end{equation}
where $\bar\rho_\mathrm{m}$ is the background density of the Universe. Considering that the total matter is composed of the dark matter and baryons, $\delta_\mathrm{m}(\mathbf{x})$ can be expressed as
\begin{equation}
	\delta_\mathrm{m}(\mathbf{x}) = f_\mathrm{dm}\delta_\mathrm{dm}(\mathbf{x})+f_\mathrm{b}\delta_\mathrm{b}(\mathbf{x}),
\end{equation}
where $f_\mathrm{dm}=(\Omega_\mathrm{m}-\Omega_\mathrm{b})/\Omega_\mathrm{m}$, and $f_\mathrm{b}=\Omega_\mathrm{b}/\Omega_\mathrm{m}$.
By convolving the total matter density field $\delta_\mathrm{m}$ with the wavelet $\Psi$, we get its CWT as
\begin{equation}
	\label{eq:cwt}
	\tilde{\delta}_\mathrm{m}(w,\mathbf{x})=\int \delta_\mathrm{m}(\mathbf{x}')\Psi(w,\mathbf{x}-\mathbf{x}')\mathrm{d}^3\mathbf{x}',
\end{equation}
in which $\Psi(w,\mathbf{x})=w^{\frac{3}{2}}\Psi(wr)$ with $r=|\mathbf{x}|$ is the rescaled isotropic wavelet at the scale of $w$. If the Fourier transform of the wavelet $\Psi(r)$ satisfies $0<|\mathcal{K}_\Psi=\int_0^{+\infty}\hat\Psi(k)/k\mathrm{d}k|<+\infty$, the density field can be recovered via
\begin{equation}
	\label{eq:inverse_cwt}
	\delta_\mathrm{m}(\mathbf{x}) = \frac{1}{\mathcal{K}_\Psi}\int_0^{+\infty}w^{\frac{1}{2}}\tilde{\delta}_\mathrm{m}(w,\mathbf{x})\mathrm{d}w.
\end{equation}
Then we define the env-WPS as 
\begin{equation}
	\label{eq:env_wps}
	\tilde{P}_\mathrm{m}(w,\delta) \equiv \langle|\tilde{\delta}_\mathrm{m}(w,\mathbf{x})|^2\rangle_{\delta_\mathrm{m}(\mathbf{x})=\delta},
\end{equation}
which is the statistical average of the squared modulus of the CWT over coordinates with the same local matter density at the scale of $w$. By averaging over all densities, the env-WPS will degenerate to the global wavelet power spectrum (global-WPS):
\begin{equation}
	\label{eq:global_wps}
	\tilde{P}_\mathrm{m}(w) = \langle|\tilde{\delta}_\mathrm{m}(w,\mathbf{x})|^2\rangle_{\text{all}\ \delta} = \langle|\tilde{\delta}_\mathrm{m}(w,\mathbf{x})|^2\rangle_\mathrm{V},
\end{equation}
where $\langle\ldots\rangle_\mathrm{V}$ denotes the statistical average over the full space. Obviously, the relationship between the global-WPS $\tilde{P}_\mathrm{m}(w)$ and the env-WPS $\tilde{P}_\mathrm{m}(w,\delta)$ is 
\begin{equation}
	\label{eq:env_global_wps}
	\tilde{P}_\mathrm{m}(w)=\sum_\delta f_\mathrm{V}(\delta)\tilde{P}_\mathrm{m}(w,\delta),
\end{equation}
where $f_V(\delta)=V_\delta/V$ is the volume fraction of the density environment of $\delta_\mathrm{m}(\mathbf{x})=\delta$. For more details of the wavelet analysis methods we developed, we refer the reader to \citet{Wang2021, Wang2022b, Wang2023}. 

To isolate the impact of baryonic processes on the matter distribution, we can compare the same statistic between the hydrodynamical simulations and their matched DMO runs with the same initial conditions. We primarily employ the env-WPS to accomplish the objective of this study, while treating others as supplementary statistics.

\subsection{Numerical procedure}

In this subsection, we present the numerical procedure for estimating the env-WPS from simulations in four steps:\footnote{With the support of \texttt{nbodykit} \citep{Hand2018}, we have developed a Python module to implement this algorithm, which is released on \url{https://github.com/WangYun1995/WPSmesh}.}
\begin{enumerate}[leftmargin=\parindent,align=left,labelwidth=\parindent,labelsep=0pt]
	\item\ \textit{Assign mass to the regular grid}
	
	To conduct our analysis, the initial step involves assigning the mass distribution of particles to the regular Cartesian grid. The mass assignment schemes usually employed include Cloud-in-Cell (CIC), Triangular-Shaped Cloud (TSC), Piecewise Cubic Spline (PCS), and the scale functions of Daubechies wavelets \citep[see][]{Cui2008,Sefusatti2016}. However, despite the last scheme is well-suited for accurately measuring Fourier-based statistics, e.g. power spectrum and bispectrum, the fields in real space obtained using it suffer from spurious artifacts. Here, we assign mass to a grid with $N_\mathrm{g}=1536^3$ cells using the PCS assignment scheme, which performs better than the CIC and TSC in underdense regions. Readers are referred to appendix \ref{sec:assign_schemes} for the tests of different schemes.

    \begin{table*}
    	\centering
    	\caption{The local density environments.}
    	\label{tab:dens_envs}
    	\noindent
    	\resizebox{\textwidth}{!}{
    		\begin{tabular}{@{}rcccccccc}
    			\hline
    			& $\delta_0$ & $\delta_1$ & $\delta_2$ & $\delta_3$ & $\delta_4$ & $\delta_5$ & $\delta_6$ & $\delta_7$ \Tstrut\\ \\[-1.2em]
    			\hline
    			$\delta_\mathrm{m}\in$   & $[-1,-0.9)$ & $[-0.9,-0.684)$  & $[-0.684,0)$ & $[0,2.162)$  & $[2.162,9)$  & $[9,30.623)$ & $[30.623,99)$ & $[99,+\infty)$ \Tstrut\\ \\[-0.5em]
    			$\rho_\mathrm{m}/\bar\rho_\mathrm{m}\in$    & $[0,0.1)$ & $[0.1,0.316)$ & $[0.316,1)$  & $[1,3.162)$ & $[3.162,10)$ & $[10,31.623)$ & $[31.623,100)$ & $[100,+\infty)$ \Bstrut\\
    			\hline
    		\end{tabular}
    	}
    \end{table*}

	\item\ \textit{Choose a specific wavelet function}
	
	In the context of the CWT, there is a wide variety of candidate wavelets. The current study requires that we focus on both the real space and the scale domain, which suggests that a wavelet with a good trade-off between spatial and scale resolutions is needed. Thus, consistent with \citet{Wang2022b}, we choose the isotropic cosine-weighted Gaussian-derived wavelet (CW-GDW) as the analyzing wavelet, the formula of which is 
	\begin{equation}
		\label{eq:iso_cwgdw}
		\Psi(\mathbf{x})=C_\mathrm{N}\left[ (4-r^2)\cos r+2(\frac{1}{r}-r)\sin r\right]e^{-\frac{r^2}{2}},
	\end{equation}
    where $r=|\mathbf{x}|$, $C_\mathrm{N}=\frac{2}{\pi^{3/4}}\sqrt{2e/(9+55e)}$ is the normalization constant such that $\int |\Psi(\mathbf{x})|^2\mathrm{d}^3\mathbf{x}=1$, and its Fourier transform is
    \begin{equation}
    	\label{eq:iso_cwgdw_k}
    	\hat\Psi(\mathbf{k})=(2\pi)^{\frac{3}{2}}C_\mathrm{N}k(k\cosh k-\sinh k)e^{-\frac{1}{2}(1+k^2)},
    \end{equation}
    where $k=|\mathbf{k}|$. For this wavelet, the scale $w$ can be expressed as the pseudo wavenumber $k_\mathrm{pseu}$ by the following relation \citep[see][]{Meyers1993,Torrence1998,Wang2022a},
    \begin{equation}
    	\label{eq:k_pseu}
    	w=c_wk_\mathrm{pseu},
    \end{equation}
    with $c_w\approx 0.3883$. The radial view of the isotropic CW-GDW is shown in Fig. \ref{fig:iso_cw_gdw}. As can be seen, it is well localized in both the real and Fourier domains.
    
	\item\ \textit{Perform the CWT by FFT}
	
	The cosmic density fields of simulations are assumed to satisfy the periodic boundary conditions, i.e. $\delta_\mathrm{m}(\mathbf{x})=\delta_\mathrm{m}(\mathbf{x}+\mathbf{n}L_\mathrm{box})$, where $\mathbf{n}=(n_x,n_y,n_z)$ is an integer vector. Therefore, equation \eqref{eq:cwt} can be rearranged as
	\begin{equation}
		\label{eq:cwt_periodic}
		\tilde{\delta}_\mathrm{m}(w,\mathbf{x})=\int_{V_\mathrm{box}}\delta_\mathrm{m}(\mathbf{x}')\Psi^P(w,\mathbf{x}-\mathbf{x}')\mathrm{d}^3\mathbf{x}',
	\end{equation}
    where $V_\mathrm{box}=L_\mathrm{box}^3$ is the volume of the simulation box, and $\Psi^P(w,\mathbf{x})$ is the periodized wavelet given by
    \begin{equation}
    	\Psi^P(w,\mathbf{x}) = \sum_\mathbf{n}\Psi(w,\mathbf{x}+\mathbf{n}L_\mathrm{box}).
    \end{equation}
    One can see that $\tilde{\delta}_\mathrm{m}(w,\mathbf{x})$ is also periodic, which thus can be expressed in terms of Fourier series as follows
    \begin{equation}
    	\tilde{\delta}_\mathrm{m}(w,\mathbf{x})=\frac{1}{V_\mathrm{box}}\sum_\mathbf{k}\hat\delta_\mathrm{m}(\mathbf{k})w^{-\frac{3}{2}}\hat\Psi\left(\frac{k}{w}\right)e^{-\mathrm{i}\mathbf{k}\cdot\mathbf{x}},
    \end{equation}
    where $\hat\delta_\mathrm{m}(\mathbf{k})$ and $w^{-\frac{3}{2}}\hat\Psi(k/w)$ are Fourier transforms of the density field and the rescaled wavelet, respectively. We see that the CWT is just the inverse Fourier transform of the multiplication of $\hat\delta_\mathrm{m}(\mathbf{k})$ and $w^{-\frac{3}{2}}\hat\Psi(k/w)$, which can be implemented by the Fast Fourier Transform (FFT) efficiently. Here the CWT of each density field is computed at $N_k=25$ equally logarithmically spaced scales in 
    \begin{equation}
    	k_\mathrm{f}\leq w/c_w\leq 0.4k_\mathrm{Nyq},
    \end{equation} 
    where $k_\mathrm{f}=2\pi/L_\mathrm{box}$, and $k_\mathrm{Nyq}=N_\mathrm{g}^{\frac{1}{3}}\pi/L_\mathrm{box}$. In this scale range, the smearing and aliasing effects, as well as the shot noise are negligible (see appendix \ref{sec:assign_schemes}).

	\begin{figure}
		\centerline{\includegraphics[width=0.45\textwidth]{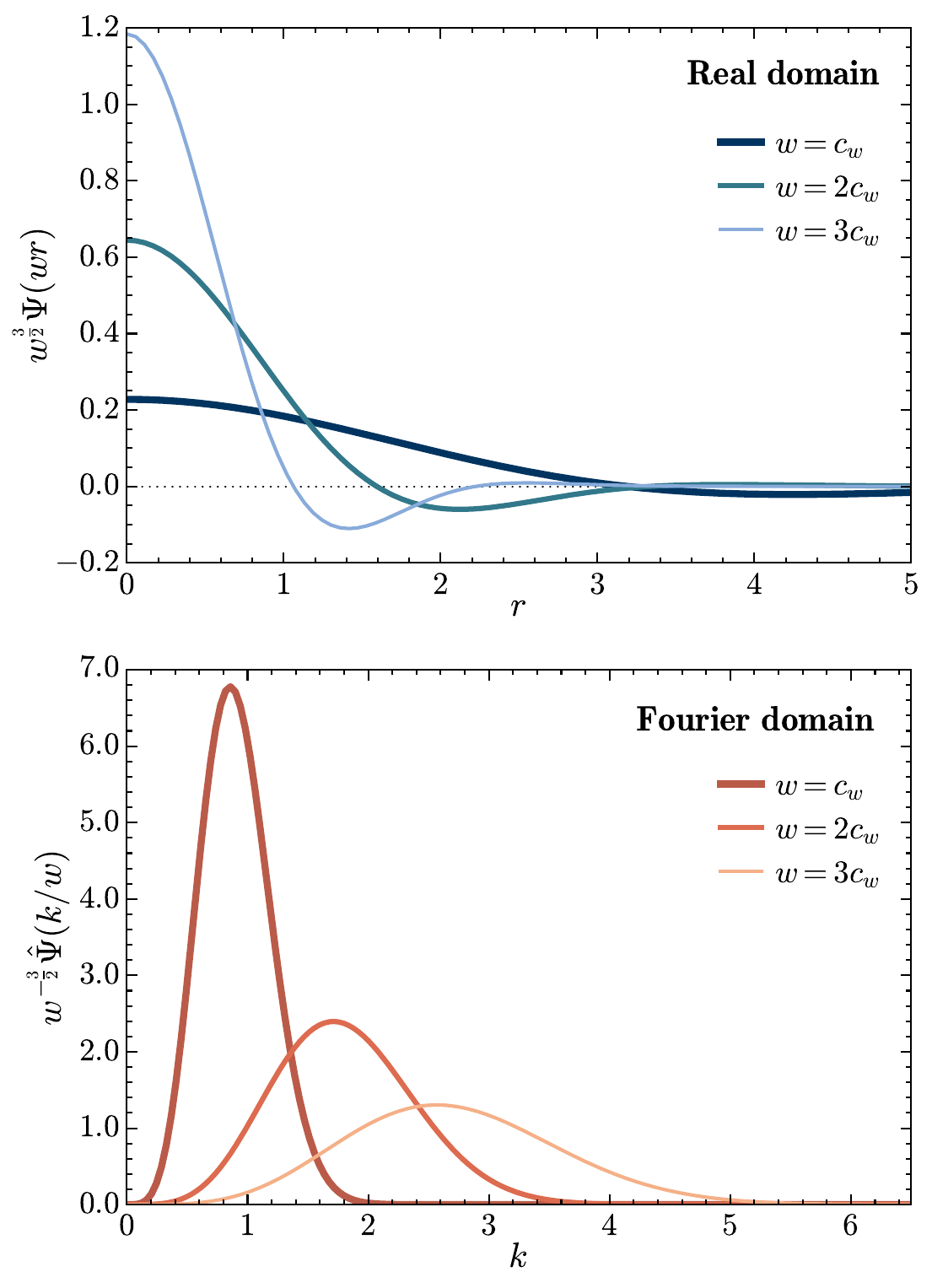}}
		\caption{The isotropic CW-GDW (top) and its Fourier transform (bottom), at the scales of $w=c_w$, $2c_w$, and $3c_w$. For presentational convenience, all
			variables are dimensionless.  }
		\label{fig:iso_cw_gdw}
	\end{figure}
 
	\item\ \textit{Measure the env-WPS}
	
	In this step, we divide the densities into $J=8$ bins: (a) six logarithmic bins equally divided
	between $\rho_\mathrm{m}/\bar\rho_\mathrm{m} = 0.1$ and $\rho_\mathrm{m}/\bar\rho_\mathrm{m} = 100$, (b) one with $\rho_\mathrm{m}/\bar\rho_\mathrm{m} < 0.1$, and (c) one with $\rho_\mathrm{m}/\bar\rho_\mathrm{m} \geq 100$, which are listed in Table \ref{tab:dens_envs}. For the $j$-th environment $\delta_j$, we compute the env-WPS with equation~(\ref{eq:env_wps}) as
	\begin{equation}
		\label{eq:measure_env_wps}
		\tilde{P}_\mathrm{m}(w,\delta_j)=\frac{1}{N_{\delta_j}}\sum_{\delta_\mathrm{m}(\mathbf{x})\in\delta_j}|\tilde{\delta}_\mathrm{m}(w,\mathbf{x})|^2,
	\end{equation}
    where $N_{\delta_j}$ is the number of cells within the environment $\delta_j$, and thus the volume fraction is given by $f_\mathrm{V}(\delta_j)=N_{\delta_j}/N_\mathrm{g}$. For the rest of the equations and plots, we will use the pseudo wavenumber $k_\mathrm{pseu}$ as a proxy for the wavelet scale $w$, and throw away its subscript ``pseu" without any ambiguity. We refer to Appendix \ref{sec:example_envwps} for a better understanding of the env-WPS.
\end{enumerate}

\begin{figure}
	\centerline{\includegraphics[width=0.48\textwidth]{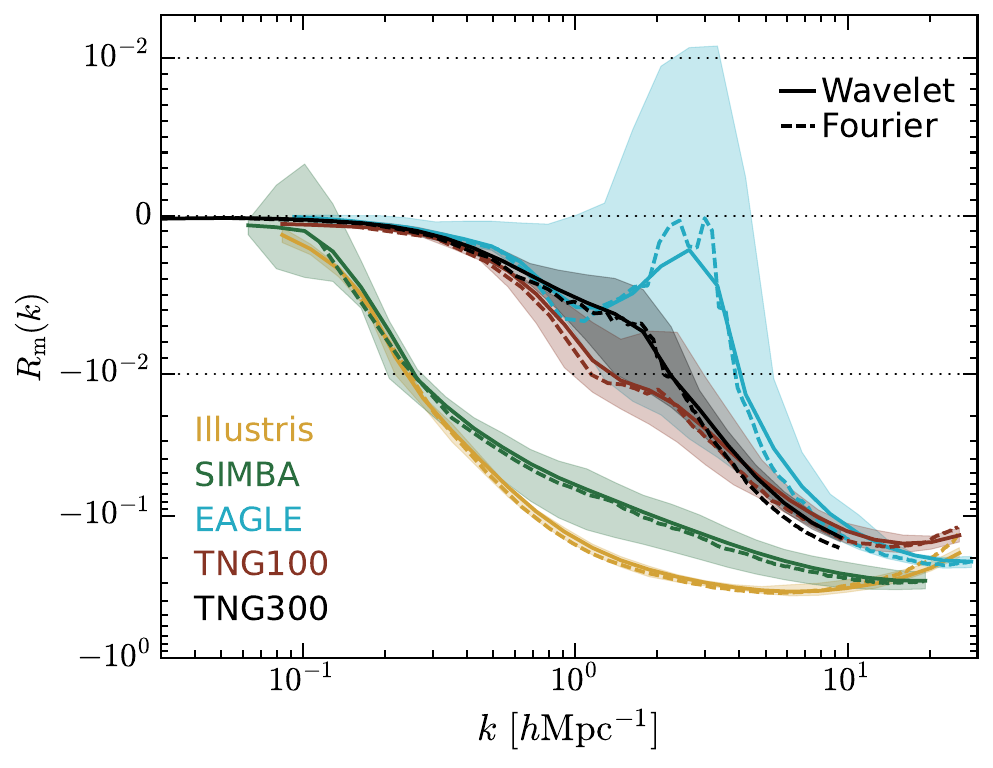}}
	\caption{A comparison of baryonic effects on the global-WPSs (solid lines) and that on the FPS (dashed lines) at $z=0$, measured from different simulations. The color bands show the cosmic variances of the relative difference $R_\mathrm{m}(k)$, estimated by the $1$-$\sigma$ dispersion of the measurements in sub-volumes.}
	\label{fig:tot_globalWPS}
\end{figure}

\begin{figure*}
	\centerline{\includegraphics[width=0.995\textwidth]{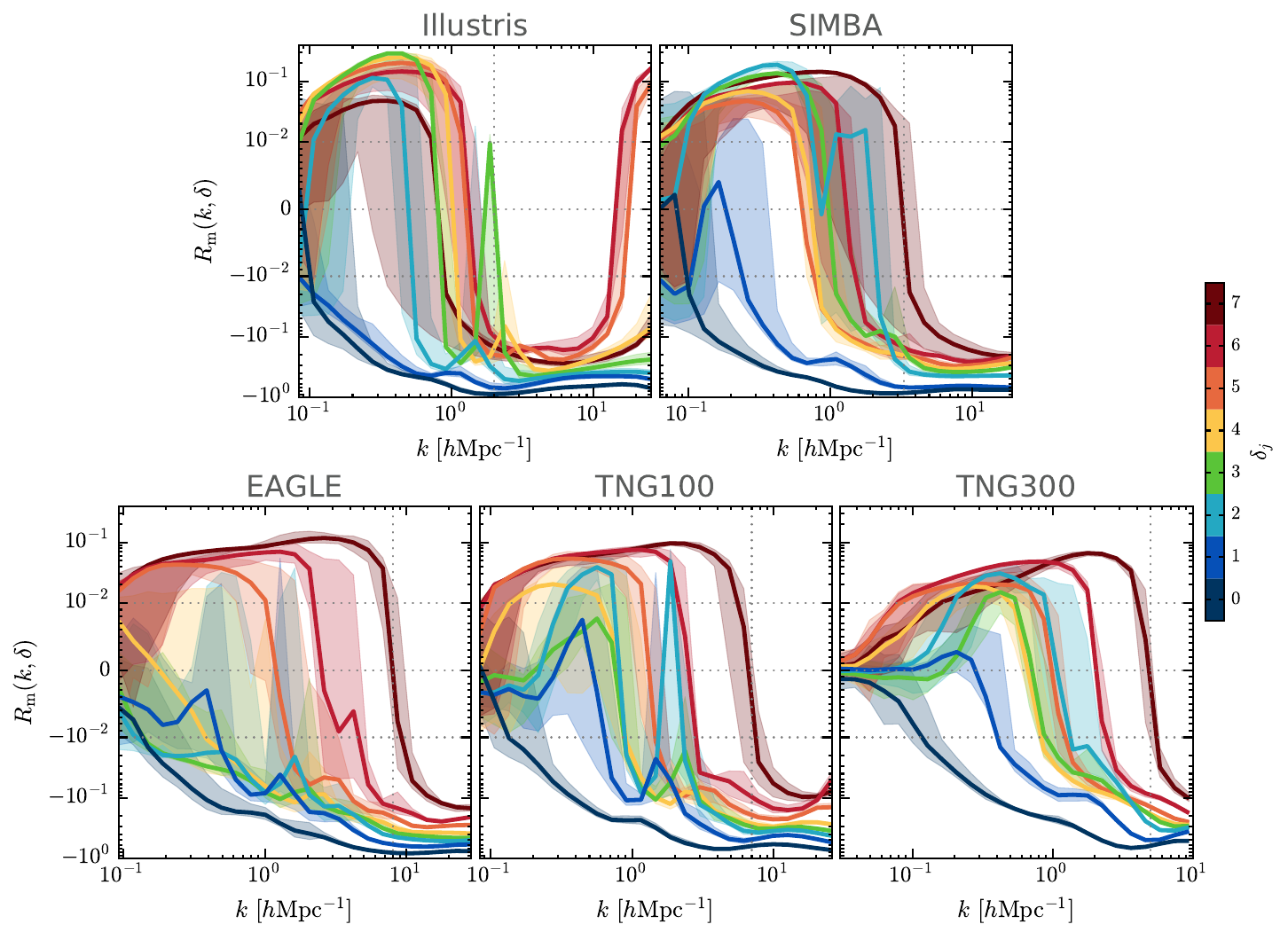}}
	\caption{The relative difference between the total matter env-WPS in hydrodynamic simulations and that in DMO runs at $z=0$, measured from Illustris, SIMBA, EAGLE, TNG100, and TNG300, respectively. As labeled by the colorbar, the colors range from dark blue to dark red, corresponding to local densities from the lowest to the highest. The light-colored areas represent the cosmic variances of $R_\mathrm{m}(k,\delta)$, quantified by the $1$-$\sigma$ dispersion of the measurements in sub-volumes. The matter clustering is suppressed in all environments when the scales are smaller than that indicated by the vertical dotted line, which is equal to $2\ h\mathrm{Mpc}^{-1}$ in Illustris, $3.3\ h\mathrm{Mpc}^{-1}$ in SIMBA, $8\ h\mathrm{Mpc}^{-1}$ in EAGLE, $7\ h\mathrm{Mpc}^{-1}$ in TNG100, and $5\ h\mathrm{Mpc}^{-1}$ in TNG300.}
	\label{fig:tot_envWPS}
\end{figure*}

\section{Baryonic effects}
\label{sec:results}

Here, we explore the baryonic effects on the total matter distribution by mainly examining the relative difference between the total matter env-WPS in hydrodynamic simulations and their corresponding DMO runs, which is given explicitly below\footnote{See Appendix \ref{sec:baryonic_eff_dm} for the relative differences measured based on the density environment of the dark matter component.}  
\begin{equation}
	R_\mathrm{m}(k,\delta) = \frac{\tilde{P}_\mathrm{m}(k,\delta)}{\tilde{P}_\mathrm{DMO}(k,\delta)} - 1.
\end{equation}
Then for the total matter global-WPS, with equation~(\ref{eq:env_global_wps}), we have
\begin{align}
\label{eq:relative_global_wps}
	R_\mathrm{m}(k) & = \frac{\tilde{P}\mathrm{m}(k)}{\tilde{P}_\mathrm{DMO}(k)} - 1 \nonumber\\
    & = \sum_\delta f_\mathrm{V}(\delta)\tilde{P}_\mathrm{m}(k,\delta)/\tilde{P}_\mathrm{DMO}(k) -1 \nonumber\\
	& = \sum_\delta[r_\mathrm{V}(\delta)+1][R_\mathrm{m}(k,\delta)+1]Q_\mathrm{DMO}(k,\delta) -1 ,
\end{align}
where $r_\mathrm{V}(\delta)+1=f_\mathrm{V}(\delta)/f^\mathrm{DMO}_\mathrm{V}(\delta)$, and $Q_\mathrm{DMO}(k,\delta)=f_\mathrm{V}^\mathrm{DMO}(\delta)\tilde{P}_\mathrm{DMO}(k,\delta)/\tilde{P}_\mathrm{DMO}(k)$ with $\sum_\delta Q_\mathrm{DMO}(k,\delta)=1$. In the above equation, the second and third equalities tell us the contribution to the relative global-WPS $\tilde{P}_\mathrm{m}(k)/\tilde{P}_\mathrm{DMO}(k)$ from individual density environment. 

Note that due to the limited volume of the simulation box and the fact that there is only one single realization of initial conditions for each simulation, the estimated global- and env-WPSs, are inevitably affected by cosmic variance. However, since the hydrodynamical run shares the same initial condition with its matched DMO run, some of the variance will cancel out in the relative differences, $R_\mathrm{m}(k)$ and $R_\mathrm{m}(k,\delta)$. To estimate the residual cosmic variance, we divide the simulation volume into eight equal-sized sub-volumes and measure the relative differences in each sub-volume separately, as is similar to the approaches in \citet{Chisari2018} and \citet{Foreman2020}. Furthermore, we quantify the spread of them by the $1$-$\sigma$ dispersion around the median values (see Appendix \ref{sec:cv}).

\subsection{Effects on the FPS and global-WPS}

The effects of baryonic physics on the FPS have been extensively investigated and discussed \citep[e.g.][]{vanDaalen2011, vanDaalen2020, Hellwing2016, Chisari2018, Springel2018, Debackere2020, Arico2020}. The consensus of these investigations is that the AGN feedback can suppress the FPS relative to the DMO simulations out to large scales of $k<1\ h\mathrm{Mpc}^{-1}$ at $z=0$. It is necessary to test the validity of our global-WPS measurements against those of the FPS, the results of which are shown in Fig. \ref{fig:tot_globalWPS}. For each simulation, the global-WPS is suppressed in the same fashion as the FPS, albeit with subtle variations. This is expected since the global-WPS is a wavelet-weighted average of the FPS over the wavenumber space \citep[see equation~(B11) in][]{Wang2022a}. Given this consistency, we do not discuss the baryonic effects on the global-WPS in detail here. 

\begin{table}
	\centering
	\caption{The maximum effects of baryonic physics observed in the extreme density environments.}
	\label{tab:Rdiff_under_over}
	\noindent
	\resizebox{0.5\textwidth}{!}{
		\begin{tabular}{@{}ccccccc}
			\hline
			&  & Illustris & SIMBA & EAGLE & TNG100 & TNG300 \Tstrut\\ \\[-1.2em]
			\hline
			\multirow{3}{*}{Suppression}   & Scale & \multirow{2}{*}{$1.86$}  & \multirow{2}{*}{$2.86$} & \multirow{2}{*}{$10.96$}  & \multirow{2}{*}{$3.81$}  &\multirow{2}{*}{$3.63$} \Tstrut\\
			& ($h\mathrm{Mpc}^{-1}$) &    &  &   &   &  \\ \\[-0.5em]
			& Amplitude & $-0.89$ & $-0.86$  & $-0.82$ & $-0.70$ & $-0.67$  \Bstrut\\
			\hline
			\multirow{3}{*}{Enhancement}   & Scale & \multirow{2}{*}{$0.35$}  & \multirow{2}{*}{$0.87$} & \multirow{2}{*}{$2.62$}  & \multirow{2}{*}{$1.86$}  & \multirow{2}{*}{$1.77$} \Tstrut\\
			& ($h\mathrm{Mpc}^{-1}$) &   &  &   &   &  \\ \\[-0.5em]
			& Amplitude & $0.05$ & $0.15$  & $0.12$ & $0.10$ &  $0.07$ \Bstrut\\
			\hline
			\multicolumn{7}{@{}p{9.8cm}}{\textit{Note.} The ``Suppression" row lists the maximum suppression and the scale where it occurs in the emptiest environment, $\delta_0$, for each simulation, while the ``Enhancement" row shows the greatest enhancement in the densest environment, $\delta_7$.}
		\end{tabular}
		
	}
\end{table}

\begin{figure}
	\centerline{\includegraphics[width=0.45\textwidth]{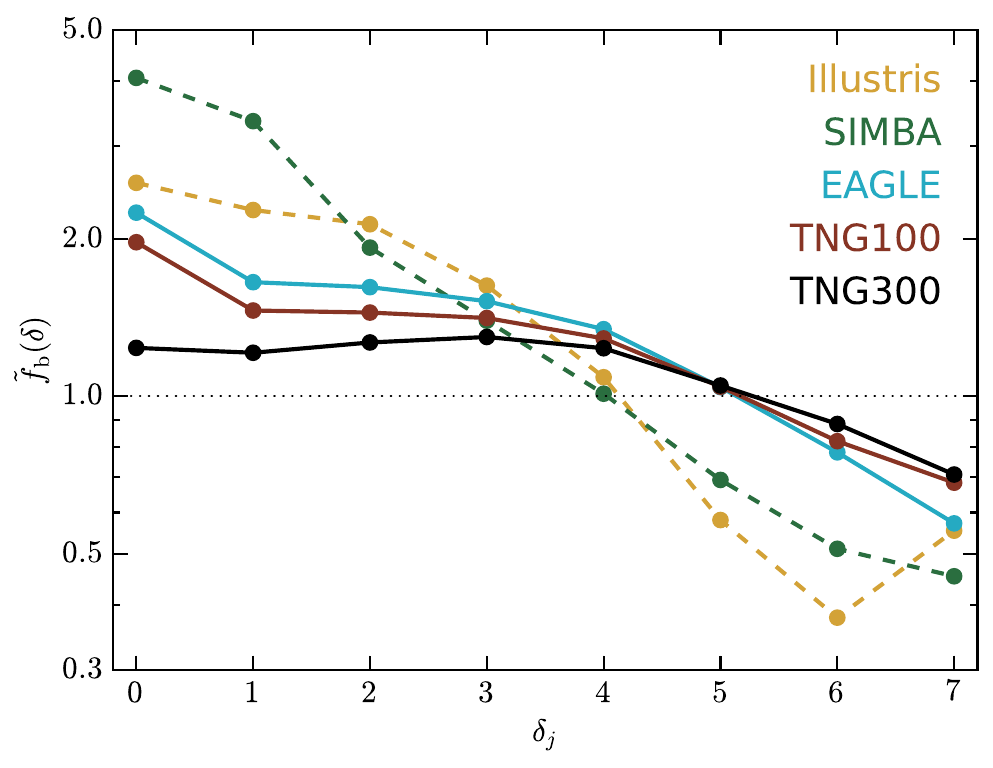}}
	\caption{The re-normalized baryon fraction $\tilde f_\mathrm{b}(\delta)$ as a function of the local density at $z=0$, measured from different simulations as labeled. We see that the baryon fraction is above the mean fraction $\Omega_\mathrm{b}/\Omega_\mathrm{m}$ at the low-density end, while below the mean at the high-density end. }
	\label{fig:env_bar_frac}
\end{figure}

\begin{figure*}
	\centerline{\includegraphics[width=0.995\textwidth]{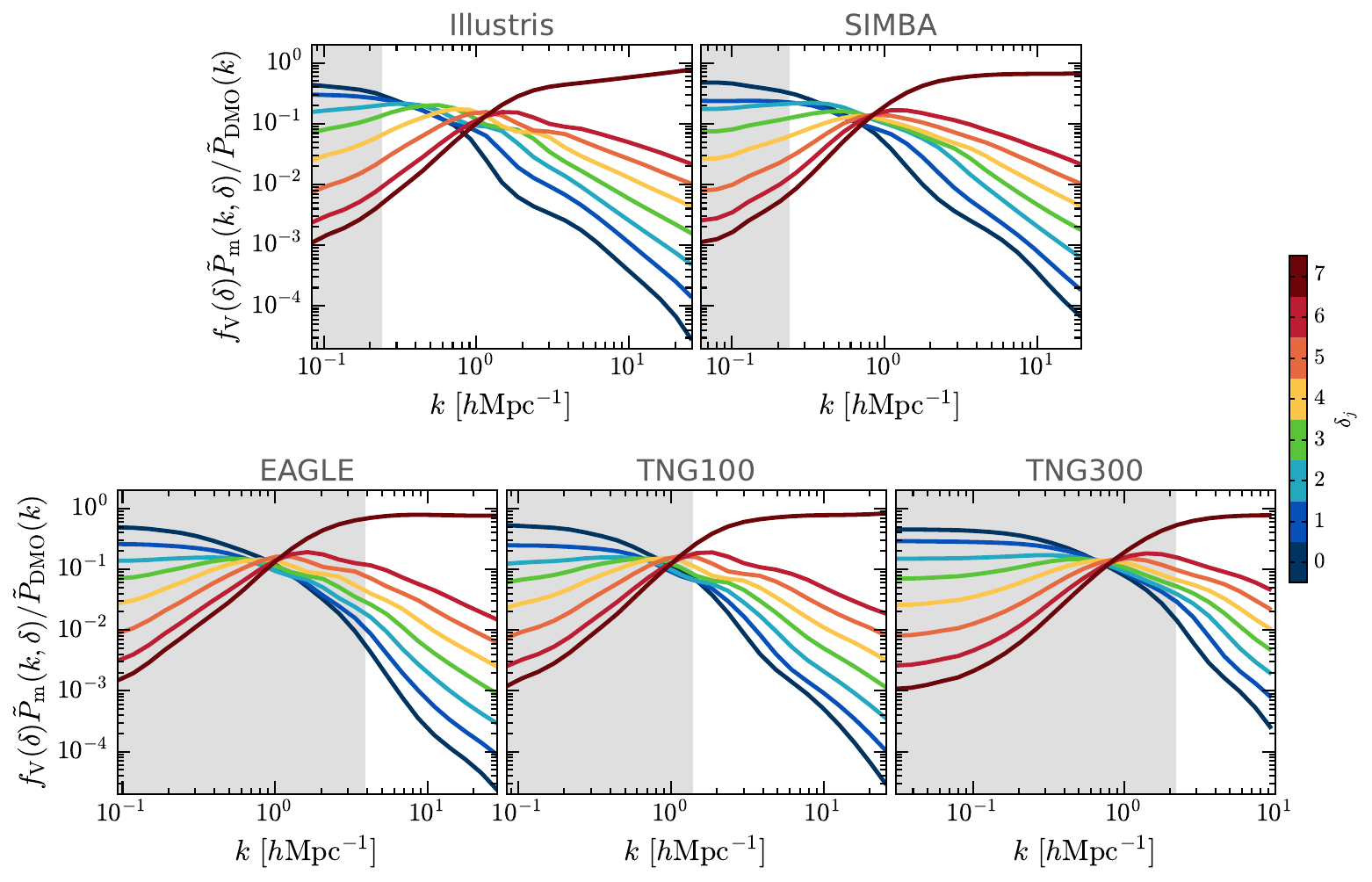}}
	\caption{The summand $f_\mathrm{V}(\delta)\tilde{P}_\mathrm{m}(k,\delta)/\tilde{P}_\mathrm{DMO}(k)$ in equation \eqref{eq:relative_global_wps}, which demonstrates the absolute contribution of each density environment to the impact of baryons on the global-WPS. In these panels, the gray shaded regions denote the scale range where the global-WPS suppression is less than 1 per cent (see Fig. \ref{fig:tot_globalWPS}).  }
	\label{fig:contribution_from_envs}
\end{figure*}

\subsection{Effects on the env-WPS}

In this subsection, we will consider how the baryonic effects on the total matter distribution may vary with scale and local density environment. Fig. \ref{fig:tot_envWPS} presents the relative difference of the total matter env-WPS in hydrodynamic runs with respect to that in DMO runs. By visual inspection, we can see that the cosmic variance has a sizable effect on large scales. A closer look shows that this effect is the minimal in TNG300, which has the largest cosmic volume among the simulations used here. However, for extremely overdense environments, the cosmic variance appears to be more severe in Illustris and SIMBA than in other simulations. Particularly, the relative differences in the environments of $\delta_{j\geqslant6}$ measured from the full volume of SIMBA, fall outside the 1-$\sigma$ uncertainty of those from eight sub-volumes, on scales of $k\lesssim 1\ h\text{Mpc}^{-1}$. This phenomenon reflects the inhomogeneity of the AGNs in SIMBA, as shown in Fig. \ref{fig:fraction_black_holes}.

Clearly, the relative difference $R_\mathrm{m}(k,\delta)$ displays a prominent scale- and environment-dependence for total matter density field. Regarding the general features, it can be seen that the env-WPS is significantly suppressed on small scales for all density environments, while is enhanced at intermediate and large scales for some environments. To be specific, the scale (labeled $k_\mathrm{th}$ for convenience) between the suppression and enhancement regimes is equal to $2\ h\mathrm{Mpc}^{-1}$ in Illustris, $3.3\ h\mathrm{Mpc}^{-1}$ in SIMBA, $8\ h\mathrm{Mpc}^{-1}$ in EAGLE, $7\ h\mathrm{Mpc}^{-1}$ in TNG100, and $5\ h\mathrm{Mpc}^{-1}$ in TNG300, respectively. On small scales of $k>k_\mathrm{th}$, the env-WPS is suppressed more strongly in lower density environments, with the exception of Illustris, in which the suppression is not monotonic with the local density. We also note that in the overdense regions of $\delta_{j\geqslant6}$, the env-WPS is enhanced by $1-10$ per cent on the galactic scales of $k> 10\ h\mathrm{Mpc}^{-1}$ for Illustris. This phenomenon may be related to more efficient star formation in Illustris \citep{Nelson2015,BeltzMohrmann2021}. Nevertheless, all simulations agree that the most significant suppression happens in the emptiest environment $\delta_0$, over almost all scales, which becomes greater than $1$ per cent on $k\sim 0.1\ h\text{Mpc}^{-1}$, and then reaches a maximum of $\sim 67-89$ per cent on scales of $1.86\lesssim k\lesssim10.96\ h\text{Mpc}^{-1}$ (see Table \ref{tab:Rdiff_under_over} for exact values). This result essentially confirms the finding of \citet{Sunseri2023}, which revealed that the probability distribution function of the density field is suppressed by nearly $100$ per cent in the emptiest regions. On the scales of $k<k_\mathrm{th}$, we see that the env-WPS is enhanced in overdense environments of $\delta_{j\geqslant5}$ (for EAGLE), as well as in several lower density environments of $\delta_{j\geqslant2}$. Among these environments, the enhancement is the most significant in the densest region at scales of $k\gtrsim1 \ h\mathrm{Mpc}^{-1}$ (except Illustris), and which peaks at scales of $0.87-2.62\ h\mathrm{Mpc}^{-1}$ with amplitudes of $7-15$ per cent (see Table \ref{tab:Rdiff_under_over}). This result is more explicit for the EAGLE, TNG100, and TNG300 simulations, in which measurements of $R_\mathrm{m}(k,\delta_7)$ are less affected by cosmic variance.

According to \citet{vanDaalen2020}, the FPS suppression due to baryonic physics can be predicted by the baryon fraction of haloes, which is indicative of the feedback strength. In our context, it is fascinating to investigate the baryon fraction of local density environments:
\begin{align}
	\label{eq:baryon_frac}
	f_\mathrm{b}(\delta_j)&=\frac{\sum_{\delta_\mathrm{m}(\mathbf{x})\in \delta_j}\rho_\mathrm{b}(\mathbf{x})\Delta V_{\rm cell}}{\sum_{\delta_\mathrm{m}(\mathbf{x})\in \delta_j}\rho_\mathrm{m}(\mathbf{x})\Delta V_{\rm cell}} \nonumber\\
	& =\frac{\Omega_\mathrm{b}}{\Omega_\mathrm{m}}\cdot\frac{\sum_{\delta_\mathrm{m}(\mathbf{x})\in \delta_j}[\delta_\mathrm{b}(\mathbf{x})+1]}{\sum_{\delta_\mathrm{m}(\mathbf{x})\in \delta_j}[\delta_\mathrm{m}(\mathbf{x})+1]},
\end{align}
in which $\Delta V_{\rm cell}=V_\mathrm{box}/N_\mathrm{g}$ is the cell volume of the simulation box. For comparison between simulations with different cosmologies, we re-normalize the fraction $f_\mathrm{b}(\delta_j)$ as
\begin{equation}
	\label{eq:renorm_bar_frac}
	\tilde f_\mathrm{b}(\delta_j) \equiv f_\mathrm{b}(\delta_j)/\frac{\Omega_\mathrm{b}}{\Omega_\mathrm{m}},
\end{equation}
the measurements of which are displayed in Fig. \ref{fig:env_bar_frac}. It is impossible here to tightly constrain the correlation between the baryon fraction and baryonic effects on the env-WPS, since we only have five simulations. Even so, these results can offer us some guidance. We find that the baryon fraction is above the mean fraction $\Omega_\mathrm{b}/\Omega_\mathrm{m}$ at the low-density end, and then decreases to below it with increasing density. This finding is attributed to the gas ejection from the very overdense regions driven by feedback processes, in which the AGN feedback plays a vital role at the redshift of $z=0$ \citep[e.g.][]{Weinberger2018, Sorini2022}. Nonetheless, we also notice that \citet{Yang2020} reported a consistent trend by examining the scatter plot of $\rho_\mathrm{b}/\rho_\mathrm{m}$ vs. $\rho_\mathrm{dm}$ (see Fig. 7 therein), in which they used the WIGEON simulation \citep{Feng2004} without incorporating star formation and any feedback. Due to the excellent performance of WIGEON in capturing shock wave and complex structures, they concluded that the gas falling into the potential well of haloes is prevented by turbulence generated from the non-linear structure formation. Turbulent heating could be a crucial mechanism that suppresses the matter power spectrum and mitigates the over-cooling problem, thereby warranting further investigation.

For Illustris and SIMBA, the baryon fraction is lower in overdense environments of $\delta_{j\geqslant 4}$, whereas higher in underdense environments of $\delta_{j\leqslant2}$ than that for other simulations. This means that AGN feedback is more efficient in Illustris and SIMBA, which can explain the heavy global-WPS suppression in these two simulations (see Fig. \ref{fig:tot_globalWPS}). Furthermore, by referring to Table \ref{tab:Rdiff_under_over} and Figs. \ref{fig:tot_envWPS} and \ref{fig:env_bar_frac}, it can be observed that in the emptiest environment, a higher baryon fraction is generally linked to a more pronounced suppression. A possible reason for this could be that the more effective AGN feedback injects more gas into the underdense regions, thereby raising gas pressure and impeding the gravitational clustering significantly. On the other hand, we see that for Illustris and SIMBA with lower baryon fractions in overdense environments of $\delta_{j\geqslant4}$, the enhancement peaks in the densest environment are shifted to larger scales, and the enhancement amplitudes in the lower densities are more significant on scales of $k\lesssim 1\ h\mathrm{Mpc}^{-1}$, compared to other simulations. Noteworthy, previous works show that the equilateral bispectrum is also enhanced on scales of $k\sim 2-3\ h\mathrm{Mpc}^{-1}$ in the hydrodynamic runs relative to that in the DMO runs (except Illustris), which is attributed to the gas reaccretion into massive haloes caused by the decreased AGN feedback strength at late-time \citep[see][]{Foreman2020,Takahashi2020,Arico2021}. The env-WPS enhancement in the environment of $\delta_7$ probably comes from the same source, as this environment largely overlaps with massive haloes.

\subsection{Contributions from different environments to the global-WPS ratio}

By examining the summand $f_\mathrm{V}(\delta)\tilde{P}_\mathrm{m}(k,\delta)/\tilde{P}_\mathrm{DMO}(k)$ in equation \eqref{eq:relative_global_wps}, we can learn about the absolute contributions of individual environments to the global-WPS ratio $R_\mathrm{m}(k)$ at different scales, which are illustrated in Fig. \ref{fig:contribution_from_envs}. Interestingly, the variation of $f_\mathrm{V}(\delta)\tilde{P}_\mathrm{m}(k,\delta)/\tilde{P}_\mathrm{DMO}(k)$ with scale and local density looks like a ``\textit{bow-tie}", which holds for all smiluations. The knots of the bow-tie are located at scales of $k\sim 0.7$, $0.9$, $0.9$, and $0.75\ h\mathrm{Mpc}^{-1}$ in the SIMBA, EAGLE, TNG100, and TNG300 simulations, respectively. However, in Illustris, the knot is much more diffuse, which covers the scales of $0.5\lesssim k\lesssim 1\ h\mathrm{Mpc}^{-1}$. At the bow-tie knot, all environments contribute equally to the global-WPS ratio. On the right side of the knot, the contribution comes mainly from the densest environment and decreases with decreasing density. This result is consistent with the previous findings, which point out that the FPS and bispectrum ratios can be interpreted well by the one-halo term on scales of $k\gtrsim 2\ h\mathrm{Mpc}^{-1}$ \citep[e.g.][]{Foreman2020}. Conversely, the densest environment contributes the least to the global-WPS ratio on the left side of the knot. For the Illustris and SIMBA simulations with more efficient AGN feedback, we note that on scales of $\sim 0.1\ h\mathrm{Mpc}^{-1}$, the few per cent suppression on the global-WPS is primarily due to underdense environments. 

\begin{figure}
	\centerline{\includegraphics[width=0.48\textwidth]{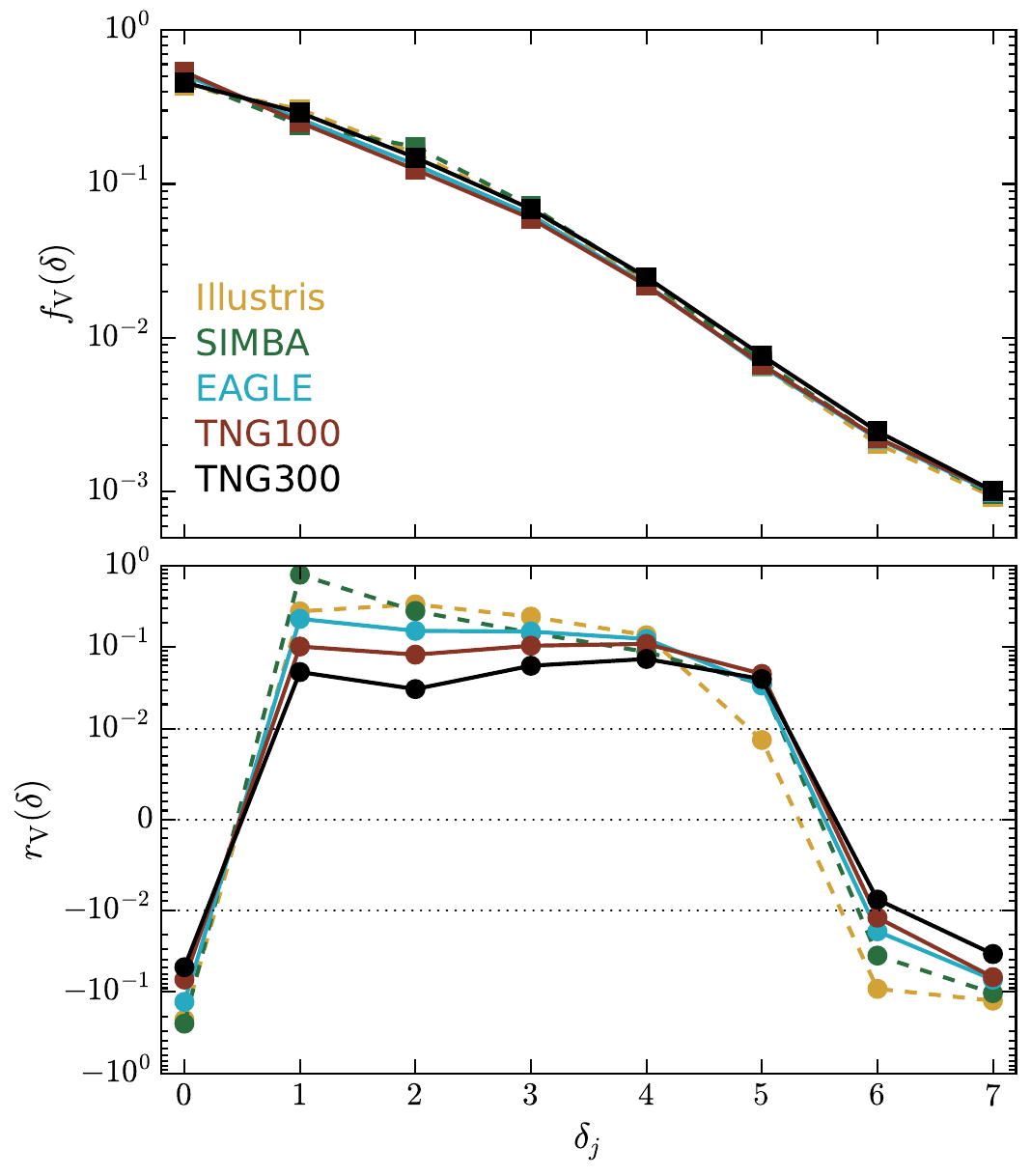}}
	\caption{\textit{Top panel}: the volume fraction occupied by each local density environment in hydrodynamic simulations. \textit{Bottom panel}: the relative difference of the environment volume fraction between the hydrodynamic and DMO runs.  }
	\label{fig:ratio_env_volume}
\end{figure}

\subsection{Volume and mass fractions of local density environments}

When we consider the relationship between the env-WPS and global-WPS, the volume fraction of the local density environment $f_\mathrm{V}(\delta)$ is an ingredient that cannot be neglected. In Fig. \ref{fig:ratio_env_volume}, we compare the volume fractions between the hydrodynamic and DMO runs. From the top panel of this figure, we see that the environment volume fractions in the different hydrodynamic simulations are approximately identical to each other, and they all decrease with increasing density. Compared to the DMO case, all the volumes of the environments are affected by $\gtrsim 1$ per cent, as shown by the bottom panel. Specifically, volumes of the extreme environments, $\delta_0$, $\delta_6$, and $\delta_7$, are reduced, while intermediate environments of $\delta_{1\leqslant j\leqslant5}$ are expanded.

\begin{figure}
	\centerline{\includegraphics[width=0.48\textwidth]{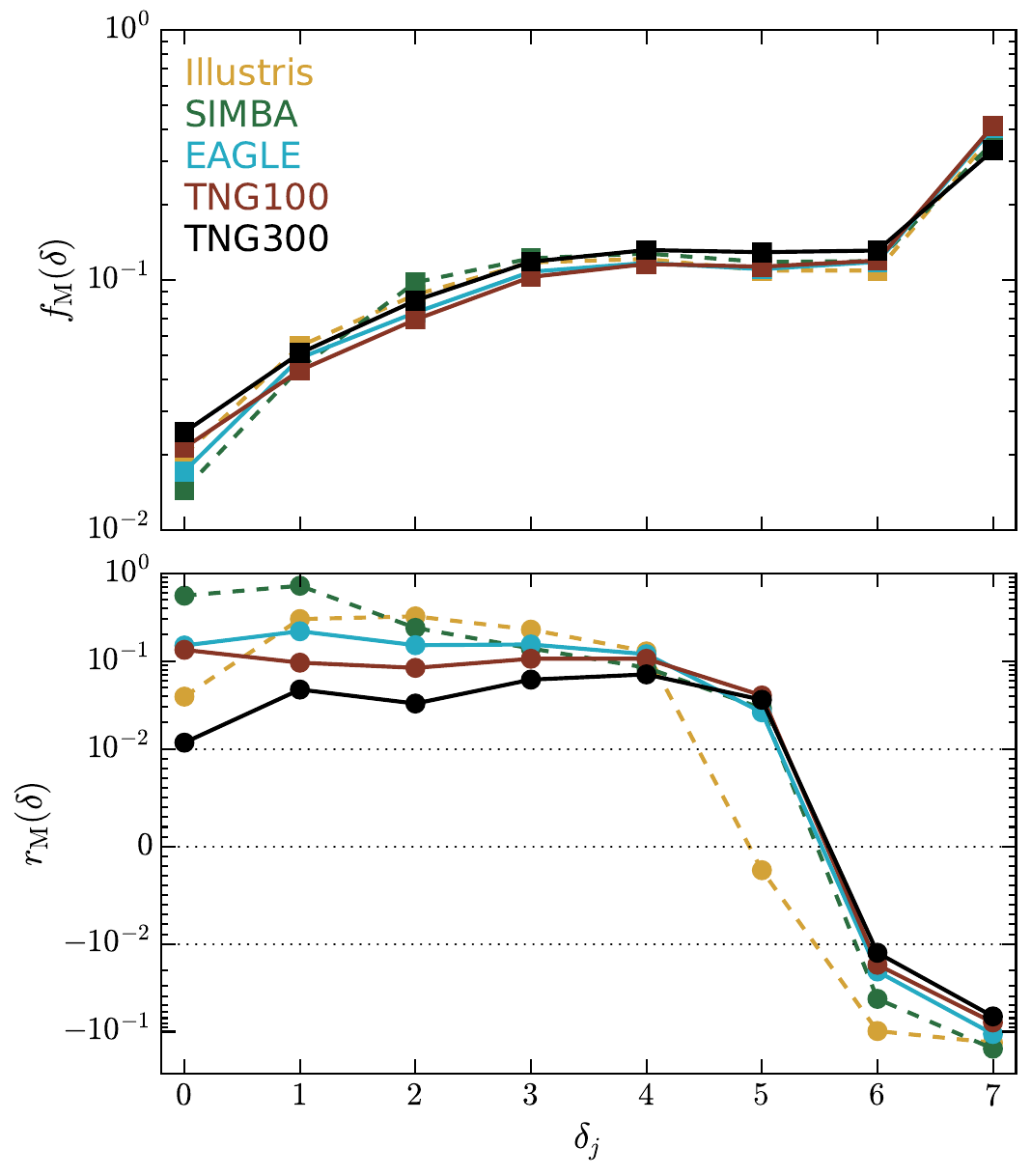}}
	\caption{\textit{Top panel}: the mass fraction of each local density environment in hydrodynamic simulations. \textit{Bottom panel}: the relative difference of the environment mass fraction between the hydrodynamic and DMO runs.  }
	\label{fig:ratio_env_mass}
\end{figure}

For completeness, we also measure the mass fractions of environments by $f_\mathrm{M}(\delta_j)=\sum_{\delta_\mathrm{m}(\mathbf{x})\in\delta_j} [\delta_\mathrm{m}(\mathbf{x})+1]/N_\mathrm{g}$, the results of which are shown in Fig. \ref{fig:ratio_env_mass}. It can be seen that in hydrodynamic simulations, the environment mass fraction grows as the local density increases. Obviously, the densest environment, which occupies only $\sim0.1$ percent of the volume, contains $\sim 40$ per cent of the mass. In contrast, with $\sim50$ per cent by volume, the emptiest environment makes a mere $\sim2$ per cent of the mass of the Universe. By examining the relative difference of the environment mass fraction with respect to the DMO runs, $r_\mathrm{M}(\delta_j)=f_\mathrm{M}(\delta_j)/f_\mathrm{M}^\mathrm{DMO}(\delta_j)-1$, the bottom panel of Fig. \ref{fig:ratio_env_mass} shows that overdense environments of $\delta_6$ and $\delta_7$ lose $\sim1-10$ per cent of their mass, while lower density environments gain a substantial mass, which is caused by the AGN feedback transporting large amounts of gas from inner halo to regions outside the halo. 

\citet{Cui2018} and \citet{Sunseri2023} also made comparisons of the mass and volume fractions of cosmic structures between the hydrodynamic and DMO runs. Using the TNG300 simulation, the latter found that haloes lose $\sim10$ percent mass to filaments, walls, and voids, while the structures change less in volume. This is in rough agreement with our results for TNG300. Surprisingly, the former supports that baryonic processes have almost no impact on large-scale structures. The sources of discrepancies may be two-fold. First, the spatial and mass resolutions of the simulations used by \citet{Cui2018} are lower than ours. Second, differences in the cosmic web classification schemes also induce differences in the results. Specifically, \citet{Cui2018} used a velocity-shear-tensor code \citep{Hoffman2012} and a tidal-tensor code \citep{Forero-Romero2009}, \citet{Sunseri2023} used the NEXUS algorithm \citep{Cautun2013}, and we divide the density field into different environments according to its face values.

Finally, taking Figs. \ref{fig:contribution_from_envs}-\ref{fig:ratio_env_mass} together, we find that the summand $f_\mathrm{V}(\delta)\tilde{P}_\mathrm{m}(k,\delta)/\tilde{P}_\mathrm{DMO}(k)$ is positively correlated with $f_\mathrm{V}(\delta)$ and anti-correlated with $f_\mathrm{M}(\delta)$ on scales larger than the bow-tie knot. Conversely, the situation is reversed on scales smaller than the bow-tie knot. This indicates that the global-WPS suppression is mainly contributed by the large-volume structures on large scales, but dominated by the massive objects on small scales.

\section{Summary and outlook}
\label{sec:summary}

The purpose of this study is to understand how the baryonic effects on the cosmic matter distribution vary with scale and local density environment. For this, we primarily use the env-WPS, defined in equation \eqref{eq:env_wps}, to interpret the present-day density fields of the total matter, which come from the SIMBA, EAGLE, Illustris, TNG100, and TNG300 simulations with volumes of $\gtrsim (100\ \mathrm{Mpc})^3$. Through a comparative analysis of the calculations between the hydrodynamic and DMO runs, we obtain some meaningful findings as follows:
\begin{enumerate}[leftmargin=\parindent,align=left,labelwidth=\parindent,labelsep=0pt]
    \item\ The matter clustering is suppressed more severely in lower density environments on small scales (except Illustris). In particular, the suppression is most severe in the emptiest environment of $\rho_\mathrm{m}/\bar\rho_\mathrm{m}<0.1$ over all scales, with a maximum of $67-89$ per cent on $1.86\lesssim k\lesssim 10.96\ h\mathrm{Mpc}^{-1}$.
	
    \item\ On intermediate and large scales, the clustering is enhanced in the environments of $\rho_\mathrm{m}/\bar\rho_\mathrm{m}\geqslant0.316$ ($\rho_\mathrm{m}/\bar\rho_\mathrm{m}\geqslant10$ in EAGLE). The most pronounced enhancement occurs in the densest environment of $\rho_\mathrm{m}/\bar\rho_\mathrm{m}\geqslant100$, which peaks on scales of $0.87\lesssim k\lesssim 2.62\ h\mathrm{Mpc}^{-1}$  with levels of $7-15$ per cent (except Illustris).
	
    \item\ The baryon fraction of the local density environment decreases with increasing density, due to the feedback processes ejecting significant amounts of gas from the highly overdense regions. Comparisons between different simulations indicate that the AGN feedback is most effective in Illustris as well as SIMBA, which could explain the main differences in the impact of baryons between them and other simulations.

    \item\ The suppression on the global-WPS is almost the same as on the Fourier power spectrum. The contributions of local density environments to the global-WPS suppression exhibit a ``bow-tie" pattern, where its knot is located around $1\ h\mathrm{Mpc}^{-1}$. On scales larger than this, the suppression mainly comes from the large-volume underdense regions, whereas from the massive objects on smaller scales.
    \item\ The mass and volume fractions of the local density environment are also affected by $\gtrsim1$ per cent due to baryon physics. In detail, the moderate-density environments of $0.1\leqslant\rho_\mathrm{m}/\bar\rho_\mathrm{m}<31.623$ are expanded, while the extreme density environments of $\rho_\mathrm{m}/\bar\rho_\mathrm{m}<0.1$ and $\rho_\mathrm{m}/\bar\rho_\mathrm{m}\geqslant31.623$ are contracted. There is a total of $\sim 10$ per cent mass transferred from the highly overdense environment of $\rho_\mathrm{m}/\bar\rho_\mathrm{m}\geqslant31.623$ to lower density environments.
\end{enumerate}

In summary, these findings support that the baryonic effects are substantial across all density environments, not limited to just the extremely overdense regions (haloes). Therefore, this study provides insights for further understanding and modeling the effects of baryonic physics. However, several issues still remain to be clarified and require further investigation, which are:
\begin{enumerate}[leftmargin=\parindent,align=left,labelwidth=\parindent,labelsep=0pt]
    \item\ The env-WPS is affected by the cosmic variance visibly on large scales and in extremely overdense environments, which suggests that the simulations with much larger volumes should be used. For example, the latest hydrodynamic simulations, MillenniumTNG \citep{Pakmor2023} and FLAMINGO \citep{Schaye2023}, are proper choices, the former with a volume of $\sim(740\ \mathrm{Mpc})^3$ and the latter with volumes of $\sim(1-2.8\ \mathrm{Gpc})^3$.
    \item\ We just qualitatively discuss the possible physical reasons underlying the results. For one thing, a more quantitative explanation requires us to employ data from multiple epochs, as well as to examine the redistribution of the baryons and the back-reaction on the dark matter. For another, different baryonic processes impact the matter distribution in different ways, which can be quantified by adjusting or disabling a certain process in the simulation, such as the star formation, black hole accretion, stellar feedback or AGN feedback.
    \item\ Predicting the baryonic effects on the env-WPS from the baryon fraction of the local density environment would require a large number of simulations. It is possible to do this with the CAMELS project \citep{VillaescusaNavarro2021}, which is a suite of $4233$ cosmological simulations with varying cosmological and feedback parameters.
    \item\ In this work, we classify the local environments only by the face value of the density field. A comparison between different classification schemes \citep[e.g.][]{Hahn2007, Forero-Romero2009, Hoffman2012, Cautun2013} is necessary in order to obtain more robust conclusions. 
\end{enumerate}

\section*{Acknowledgments}

The authors would like to acknowledge the anonymous referee for valuable comments and suggestions.
The authors also thank the Illustris, IllustrisTNG, EAGLE, and SIMBA teams for their publicly available data. 
Y.W. especially thanks Dr. Yu Feng and Dr. Simon Foreman for their helpful discussions. 
This work is supported by the National Science Foundation of China (No. 12147217, 12347163), and the Natural Science Foundation of Jilin Province, China (No. 20180101228JC).

\section*{Data Availability}
The Python codes and measurements of the env-WPS are available at \url{https://github.com/WangYun1995/WPSmesh}.

\bibliography{references}{}

\begin{thebibliography}{}
\makeatletter
\relax
\def\mn@urlcharsother{\let\do\@makeother \do\$\do\&\do\#\do\^\do\_\do\%\do\~}
\def\mn@doi{\begingroup\mn@urlcharsother \@ifnextchar [ {\mn@doi@}
  {\mn@doi@[]}}
\def\mn@doi@[#1]#2{\def\@tempa{#1}\ifx\@tempa\@empty \href
  {http://dx.doi.org/#2} {doi:#2}\else \href {http://dx.doi.org/#2} {#1}\fi
  \endgroup}
\def\mn@eprint#1#2{\mn@eprint@#1:#2::\@nil}
\def\mn@eprint@arXiv#1{\href {http://arxiv.org/abs/#1} {{\tt arXiv:#1}}}
\def\mn@eprint@dblp#1{\href {http://dblp.uni-trier.de/rec/bibtex/#1.xml}
  {dblp:#1}}
\def\mn@eprint@#1:#2:#3:#4\@nil{\def\@tempa {#1}\def\@tempb {#2}\def\@tempc
  {#3}\ifx \@tempc \@empty \let \@tempc \@tempb \let \@tempb \@tempa \fi \ifx
  \@tempb \@empty \def\@tempb {arXiv}\fi \@ifundefined
  {mn@eprint@\@tempb}{\@tempb:\@tempc}{\expandafter \expandafter \csname
  mn@eprint@\@tempb\endcsname \expandafter{\@tempc}}}

\bibitem[\protect\citeauthoryear{{Aric{\`o}}, {Angulo},
  {Hern{\'a}ndez-Monteagudo}, {Contreras}, {Zennaro}, {Pellejero-Iba{\~n}ez}
  \& {Rosas-Guevara}}{{Aric{\`o}} et~al.}{2020}]{Arico2020}
{Aric{\`o}} G.,  {Angulo} R.~E.,  {Hern{\'a}ndez-Monteagudo} C.,  {Contreras}
  S.,  {Zennaro} M.,  {Pellejero-Iba{\~n}ez} M.,   {Rosas-Guevara} Y.,  2020,
  \mn@doi [\mnras] {10.1093/mnras/staa1478}, \href
  {https://ui.adsabs.harvard.edu/abs/2020MNRAS.495.4800A} {495, 4800}

\bibitem[\protect\citeauthoryear{{Aric{\`o}}, {Angulo},
  {Hern{\'a}ndez-Monteagudo}, {Contreras}  \& {Zennaro}}{{Aric{\`o}}
  et~al.}{2021}]{Arico2021}
{Aric{\`o}} G.,  {Angulo} R.~E.,  {Hern{\'a}ndez-Monteagudo} C.,  {Contreras}
  S.,   {Zennaro} M.,  2021, \mn@doi [\mnras] {10.1093/mnras/stab699}, \href
  {https://ui.adsabs.harvard.edu/abs/2021MNRAS.503.3596A} {503, 3596}

\bibitem[\protect\citeauthoryear{{Beltz-Mohrmann} \&
  {Berlind}}{{Beltz-Mohrmann} \& {Berlind}}{2021}]{BeltzMohrmann2021}
{Beltz-Mohrmann} G.~D.,  {Berlind} A.~A.,  2021, \mn@doi [\apj]
  {10.3847/1538-4357/ac1e27}, \href
  {https://ui.adsabs.harvard.edu/abs/2021ApJ...921..112B} {921, 112}

\bibitem[\protect\citeauthoryear{{Boylan-Kolchin}, {Bullock}  \&
  {Kaplinghat}}{{Boylan-Kolchin} et~al.}{2011}]{Boylan-Kolchin2011}
{Boylan-Kolchin} M.,  {Bullock} J.~S.,   {Kaplinghat} M.,  2011, \mn@doi
  [\mnras] {10.1111/j.1745-3933.2011.01074.x}, \href
  {https://ui.adsabs.harvard.edu/abs/2011MNRAS.415L..40B} {415, L40}

\bibitem[\protect\citeauthoryear{{Bullock} \& {Boylan-Kolchin}}{{Bullock} \&
  {Boylan-Kolchin}}{2017}]{Bullock2017}
{Bullock} J.~S.,  {Boylan-Kolchin} M.,  2017, \mn@doi [\araa]
  {10.1146/annurev-astro-091916-055313}, \href
  {https://ui.adsabs.harvard.edu/abs/2017ARA&A..55..343B} {55, 343}

\bibitem[\protect\citeauthoryear{{Butsky} et~al.,}{{Butsky}
  et~al.}{2016}]{Butsky2016}
{Butsky} I.,  et~al., 2016, \mn@doi [\mnras] {10.1093/mnras/stw1688}, \href
  {https://ui.adsabs.harvard.edu/abs/2016MNRAS.462..663B} {462, 663}

\bibitem[\protect\citeauthoryear{{Castro}, {Borgani}, {Dolag}, {Marra},
  {Quartin}, {Saro}  \& {Sefusatti}}{{Castro} et~al.}{2021}]{Castro2021}
{Castro} T.,  {Borgani} S.,  {Dolag} K.,  {Marra} V.,  {Quartin} M.,  {Saro}
  A.,   {Sefusatti} E.,  2021, \mn@doi [\mnras] {10.1093/mnras/staa3473}, \href
  {https://ui.adsabs.harvard.edu/abs/2021MNRAS.500.2316C} {500, 2316}

\bibitem[\protect\citeauthoryear{{Cataldi}, {Pedrosa}, {Tissera}  \&
  {Artale}}{{Cataldi} et~al.}{2021}]{Cataldi2021}
{Cataldi} P.,  {Pedrosa} S.~E.,  {Tissera} P.~B.,   {Artale} M.~C.,  2021,
  \mn@doi [\mnras] {10.1093/mnras/staa3988}, \href
  {https://ui.adsabs.harvard.edu/abs/2021MNRAS.501.5679C} {501, 5679}

\bibitem[\protect\citeauthoryear{{Cataldi} et~al.,}{{Cataldi}
  et~al.}{2023}]{Cataldi2023}
{Cataldi} P.,  et~al., 2023, \mn@doi [\mnras] {10.1093/mnras/stad1601}, \href
  {https://ui.adsabs.harvard.edu/abs/2023MNRAS.523.1919C} {523, 1919}

\bibitem[\protect\citeauthoryear{{Cautun}, {van de Weygaert}  \&
  {Jones}}{{Cautun} et~al.}{2013}]{Cautun2013}
{Cautun} M.,  {van de Weygaert} R.,   {Jones} B. J.~T.,  2013, \mn@doi [\mnras]
  {10.1093/mnras/sts416}, \href
  {https://ui.adsabs.harvard.edu/abs/2013MNRAS.429.1286C} {429, 1286}

\bibitem[\protect\citeauthoryear{{Chan}, {Kere{\v{s}}}, {O{\~n}orbe},
  {Hopkins}, {Muratov}, {Faucher-Gigu{\`e}re}  \& {Quataert}}{{Chan}
  et~al.}{2015}]{Chan2015}
{Chan} T.~K.,  {Kere{\v{s}}} D.,  {O{\~n}orbe} J.,  {Hopkins} P.~F.,  {Muratov}
  A.~L.,  {Faucher-Gigu{\`e}re} C.~A.,   {Quataert} E.,  2015, \mn@doi [\mnras]
  {10.1093/mnras/stv2165}, \href
  {https://ui.adsabs.harvard.edu/abs/2015MNRAS.454.2981C} {454, 2981}

\bibitem[\protect\citeauthoryear{{Chisari} et~al.,}{{Chisari}
  et~al.}{2018}]{Chisari2018}
{Chisari} N.~E.,  et~al., 2018, \mn@doi [\mnras] {10.1093/mnras/sty2093}, \href
  {https://ui.adsabs.harvard.edu/abs/2018MNRAS.480.3962C} {480, 3962}

\bibitem[\protect\citeauthoryear{{Chua}, {Pillepich}, {Vogelsberger}  \&
  {Hernquist}}{{Chua} et~al.}{2019}]{Chua2019}
{Chua} K. T.~E.,  {Pillepich} A.,  {Vogelsberger} M.,   {Hernquist} L.,  2019,
  \mn@doi [\mnras] {10.1093/mnras/sty3531}, \href
  {https://ui.adsabs.harvard.edu/abs/2019MNRAS.484..476C} {484, 476}

\bibitem[\protect\citeauthoryear{{Chua}, {Vogelsberger}, {Pillepich}  \&
  {Hernquist}}{{Chua} et~al.}{2022}]{Chua2022}
{Chua} K. T.~E.,  {Vogelsberger} M.,  {Pillepich} A.,   {Hernquist} L.,  2022,
  \mn@doi [\mnras] {10.1093/mnras/stac1897}, \href
  {https://ui.adsabs.harvard.edu/abs/2022MNRAS.515.2681C} {515, 2681}

\bibitem[\protect\citeauthoryear{Crain \& van~de Voort}{Crain \& van~de
  Voort}{2023}]{Crain2023}
Crain R.~A.,  van~de Voort F.,  2023, \mn@doi [\araa]
  {10.1146/annurev-astro-041923-043618}, 61, 473

\bibitem[\protect\citeauthoryear{{Crain} et~al.,}{{Crain}
  et~al.}{2015}]{Crain2015}
{Crain} R.~A.,  et~al., 2015, \mn@doi [\mnras] {10.1093/mnras/stv725}, \href
  {https://ui.adsabs.harvard.edu/abs/2015MNRAS.450.1937C} {450, 1937}

\bibitem[\protect\citeauthoryear{{Cui}, {Liu}, {Yang}, {Wang}, {Feng}  \&
  {Springel}}{{Cui} et~al.}{2008}]{Cui2008}
{Cui} W.,  {Liu} L.,  {Yang} X.,  {Wang} Y.,  {Feng} L.,   {Springel} V.,
  2008, \mn@doi [\apj] {10.1086/592079}, \href
  {https://ui.adsabs.harvard.edu/abs/2008ApJ...687..738C} {687, 738}

\bibitem[\protect\citeauthoryear{{Cui}, {Borgani}  \& {Murante}}{{Cui}
  et~al.}{2014}]{Cui2014}
{Cui} W.,  {Borgani} S.,   {Murante} G.,  2014, \mn@doi [\mnras]
  {10.1093/mnras/stu673}, \href
  {https://ui.adsabs.harvard.edu/abs/2014MNRAS.441.1769C} {441, 1769}

\bibitem[\protect\citeauthoryear{{Cui}, {Knebe}, {Yepes}, {Yang}, {Borgani},
  {Kang}, {Power}  \& {Staveley-Smith}}{{Cui} et~al.}{2018}]{Cui2018}
{Cui} W.,  {Knebe} A.,  {Yepes} G.,  {Yang} X.,  {Borgani} S.,  {Kang} X.,
  {Power} C.,   {Staveley-Smith} L.,  2018, \mn@doi [\mnras]
  {10.1093/mnras/stx2323}, \href
  {https://ui.adsabs.harvard.edu/abs/2018MNRAS.473...68C} {473, 68}

\bibitem[\protect\citeauthoryear{Daubechies}{Daubechies}{1992}]{Daubechies1992}
Daubechies I.,  1992, Ten lectures on wavelets.
SIAM

\bibitem[\protect\citeauthoryear{Davé, Anglés-Alcázar, Narayanan, Li,
  Rafieferantsoa  \& Appleby}{Davé et~al.}{2019}]{Dave2019}
Davé R.,  Anglés-Alcázar D.,  Narayanan D.,  Li Q.,  Rafieferantsoa M.~H.,
  Appleby S.,  2019, \mn@doi [\mnras] {10.1093/mnras/stz937}, 486, 2827

\bibitem[\protect\citeauthoryear{{Debackere}, {Schaye}  \&
  {Hoekstra}}{{Debackere} et~al.}{2020}]{Debackere2020}
{Debackere} S. N.~B.,  {Schaye} J.,   {Hoekstra} H.,  2020, \mn@doi [\mnras]
  {10.1093/mnras/stz3446}, \href
  {https://ui.adsabs.harvard.edu/abs/2020MNRAS.492.2285D} {492, 2285}

\bibitem[\protect\citeauthoryear{{Engler} et~al.,}{{Engler}
  et~al.}{2021}]{Engler2021}
{Engler} C.,  et~al., 2021, \mn@doi [\mnras] {10.1093/mnras/stab2437}, \href
  {https://ui.adsabs.harvard.edu/abs/2021MNRAS.507.4211E} {507, 4211}

\bibitem[\protect\citeauthoryear{{Feng}, {Shu}  \& {Zhang}}{{Feng}
  et~al.}{2004}]{Feng2004}
{Feng} L.-L.,  {Shu} C.-W.,   {Zhang} M.,  2004, \mn@doi [\apj]
  {10.1086/422513}, \href
  {https://ui.adsabs.harvard.edu/abs/2004ApJ...612....1F} {612, 1}

\bibitem[\protect\citeauthoryear{{Foreman}, {Coulton}, {Villaescusa-Navarro}
  \& {Barreira}}{{Foreman} et~al.}{2020}]{Foreman2020}
{Foreman} S.,  {Coulton} W.,  {Villaescusa-Navarro} F.,   {Barreira} A.,  2020,
  \mn@doi [\mnras] {10.1093/mnras/staa2523}, \href
  {https://ui.adsabs.harvard.edu/abs/2020MNRAS.498.2887F} {498, 2887}

\bibitem[\protect\citeauthoryear{{Forero-Romero}, {Hoffman}, {Gottl{\"o}ber},
  {Klypin}  \& {Yepes}}{{Forero-Romero} et~al.}{2009}]{Forero-Romero2009}
{Forero-Romero} J.~E.,  {Hoffman} Y.,  {Gottl{\"o}ber} S.,  {Klypin} A.,
  {Yepes} G.,  2009, \mn@doi [\mnras] {10.1111/j.1365-2966.2009.14885.x}, \href
  {https://ui.adsabs.harvard.edu/abs/2009MNRAS.396.1815F} {396, 1815}

\bibitem[\protect\citeauthoryear{{Frenk} \& {White}}{{Frenk} \&
  {White}}{2012}]{Frenk2012}
{Frenk} C.~S.,  {White} S.~D.~M.,  2012, \mn@doi [Annalen der Physik]
  {10.1002/andp.201200212}, \href
  {https://ui.adsabs.harvard.edu/abs/2012AnP...524..507F} {524, 507}

\bibitem[\protect\citeauthoryear{Genel et~al.,}{Genel et~al.}{2014}]{Genel2014}
Genel S.,  et~al., 2014, \mn@doi [\mnras] {10.1093/mnras/stu1654}, 445, 175

\bibitem[\protect\citeauthoryear{{Hahn}, {Carollo}, {Porciani}  \&
  {Dekel}}{{Hahn} et~al.}{2007}]{Hahn2007}
{Hahn} O.,  {Carollo} C.~M.,  {Porciani} C.,   {Dekel} A.,  2007, \mn@doi
  [\mnras] {10.1111/j.1365-2966.2007.12249.x}, \href
  {https://ui.adsabs.harvard.edu/abs/2007MNRAS.381...41H} {381, 41}

\bibitem[\protect\citeauthoryear{{Hand}, {Feng}, {Beutler}, {Li}, {Modi},
  {Seljak}  \& {Slepian}}{{Hand} et~al.}{2018}]{Hand2018}
{Hand} N.,  {Feng} Y.,  {Beutler} F.,  {Li} Y.,  {Modi} C.,  {Seljak} U.,
  {Slepian} Z.,  2018, \mn@doi [\aj] {10.3847/1538-3881/aadae0}, \href
  {https://ui.adsabs.harvard.edu/abs/2018AJ....156..160H} {156, 160}

\bibitem[\protect\citeauthoryear{{Hellwing}, {Schaller}, {Frenk}, {Theuns},
  {Schaye}, {Bower}  \& {Crain}}{{Hellwing} et~al.}{2016}]{Hellwing2016}
{Hellwing} W.~A.,  {Schaller} M.,  {Frenk} C.~S.,  {Theuns} T.,  {Schaye} J.,
  {Bower} R.~G.,   {Crain} R.~A.,  2016, \mn@doi [\mnras]
  {10.1093/mnrasl/slw081}, \href
  {https://ui.adsabs.harvard.edu/abs/2016MNRAS.461L..11H} {461, L11}

\bibitem[\protect\citeauthoryear{{Henson}, {Barnes}, {Kay}, {McCarthy}  \&
  {Schaye}}{{Henson} et~al.}{2017}]{Henson2017}
{Henson} M.~A.,  {Barnes} D.~J.,  {Kay} S.~T.,  {McCarthy} I.~G.,   {Schaye}
  J.,  2017, \mn@doi [\mnras] {10.1093/mnras/stw2899}, \href
  {https://ui.adsabs.harvard.edu/abs/2017MNRAS.465.3361H} {465, 3361}

\bibitem[\protect\citeauthoryear{{Hinshaw} et~al.,}{{Hinshaw}
  et~al.}{2013}]{Hinshaw2013}
{Hinshaw} G.,  et~al., 2013, \mn@doi [\apjs] {10.1088/0067-0049/208/2/19},
  \href {https://ui.adsabs.harvard.edu/abs/2013ApJS..208...19H} {208, 19}

\bibitem[\protect\citeauthoryear{{Hoffman}, {Metuki}, {Yepes}, {Gottl{\"o}ber},
  {Forero-Romero}, {Libeskind}  \& {Knebe}}{{Hoffman}
  et~al.}{2012}]{Hoffman2012}
{Hoffman} Y.,  {Metuki} O.,  {Yepes} G.,  {Gottl{\"o}ber} S.,  {Forero-Romero}
  J.~E.,  {Libeskind} N.~I.,   {Knebe} A.,  2012, \mn@doi [\mnras]
  {10.1111/j.1365-2966.2012.21553.x}, \href
  {https://ui.adsabs.harvard.edu/abs/2012MNRAS.425.2049H} {425, 2049}

\bibitem[\protect\citeauthoryear{{Hopkins}}{{Hopkins}}{2015}]{Hopkins2015}
{Hopkins} P.~F.,  2015, \mn@doi [\mnras] {10.1093/mnras/stv195}, \href
  {https://ui.adsabs.harvard.edu/abs/2015MNRAS.450...53H} {450, 53}

\bibitem[\protect\citeauthoryear{Klypin, Kravtsov, Valenzuela  \& Prada}{Klypin
  et~al.}{1999}]{Klypin1999}
Klypin A.,  Kravtsov A.~V.,  Valenzuela O.,   Prada F.,  1999, \mn@doi [\apj]
  {10.1086/307643}, 522, 82

\bibitem[\protect\citeauthoryear{Marinacci et~al.,}{Marinacci
  et~al.}{2018}]{Marinacci2018}
Marinacci F.,  et~al., 2018, \mn@doi [\mnras] {10.1093/mnras/sty2206}, 480,
  5113

\bibitem[\protect\citeauthoryear{{McCarthy} et~al.,}{{McCarthy}
  et~al.}{2010}]{McCarthy2010}
{McCarthy} I.~G.,  et~al., 2010, \mn@doi [\mnras]
  {10.1111/j.1365-2966.2010.16750.x}, \href
  {https://ui.adsabs.harvard.edu/abs/2010MNRAS.406..822M} {406, 822}

\bibitem[\protect\citeauthoryear{{McCarthy}, {Schaye}, {Bower}, {Ponman},
  {Booth}, {Dalla Vecchia}  \& {Springel}}{{McCarthy}
  et~al.}{2011}]{McCarthy2011}
{McCarthy} I.~G.,  {Schaye} J.,  {Bower} R.~G.,  {Ponman} T.~J.,  {Booth}
  C.~M.,  {Dalla Vecchia} C.,   {Springel} V.,  2011, \mn@doi [\mnras]
  {10.1111/j.1365-2966.2010.18033.x}, \href
  {https://ui.adsabs.harvard.edu/abs/2011MNRAS.412.1965M} {412, 1965}

\bibitem[\protect\citeauthoryear{Meyers, Kelly  \& O'Brien}{Meyers
  et~al.}{1993}]{Meyers1993}
Meyers S.~D.,  Kelly B.~G.,   O'Brien J.~J.,  1993, \mn@doi [MWR]
  {10.1175/1520-0493(1993)121<2858:AITWAI>2.0.CO;2}, 121, 2858

\bibitem[\protect\citeauthoryear{Naiman et~al.,}{Naiman
  et~al.}{2018}]{Naiman2018}
Naiman J.~P.,  et~al., 2018, \mn@doi [\mnras] {10.1093/mnras/sty618}, 477, 1206

\bibitem[\protect\citeauthoryear{Nelson et~al.,}{Nelson
  et~al.}{2015}]{Nelson2015}
Nelson D.,  et~al., 2015, \mn@doi [Astron. Comput]
  {10.1016/j.ascom.2015.09.003}, 13, 12

\bibitem[\protect\citeauthoryear{Nelson et~al.,}{Nelson
  et~al.}{2018}]{Nelson2018}
Nelson D.,  et~al., 2018, \mn@doi [\mnras] {10.1093/mnras/stx3040}, 475, 624

\bibitem[\protect\citeauthoryear{Nelson et~al.,}{Nelson
  et~al.}{2019}]{Nelson2019}
Nelson D.,  et~al., 2019, \mn@doi [CompAC] {10.1186/s40668-019-0028-x}, 6, 1

\bibitem[\protect\citeauthoryear{{Oman} et~al.,}{{Oman}
  et~al.}{2015}]{Oman2015}
{Oman} K.~A.,  et~al., 2015, \mn@doi [\mnras] {10.1093/mnras/stv1504}, \href
  {https://ui.adsabs.harvard.edu/abs/2015MNRAS.452.3650O} {452, 3650}

\bibitem[\protect\citeauthoryear{{Pakmor} et~al.,}{{Pakmor}
  et~al.}{2023}]{Pakmor2023}
{Pakmor} R.,  et~al., 2023, \mn@doi [\mnras] {10.1093/mnras/stac3620}, \href
  {https://ui.adsabs.harvard.edu/abs/2023MNRAS.524.2539P} {524, 2539}

\bibitem[\protect\citeauthoryear{Pillepich et~al.,}{Pillepich
  et~al.}{2018a}]{Pillepich2018a}
Pillepich A.,  et~al., 2018a, \mn@doi [\mnras] {10.1093/mnras/stx2656}, 473,
  4077

\bibitem[\protect\citeauthoryear{Pillepich et~al.,}{Pillepich
  et~al.}{2018b}]{Pillepich2018b}
Pillepich A.,  et~al., 2018b, \mn@doi [\mnras] {10.1093/mnras/stx3112}, 475,
  648

\bibitem[\protect\citeauthoryear{{Planck Collaboration XIII}}{{Planck
  Collaboration XIII}}{2016}]{Planck2015}
{Planck Collaboration XIII} 2016, \mn@doi [\aap] {10.1051/0004-6361/201525830},
  \href {https://ui.adsabs.harvard.edu/abs/2016A&A...594A..13P} {594, A13}

\bibitem[\protect\citeauthoryear{{Planck Collaboration XVI}}{{Planck
  Collaboration XVI}}{2014}]{Planck2013}
{Planck Collaboration XVI} 2014, \mn@doi [\aap] {10.1051/0004-6361/201321591},
  \href {https://ui.adsabs.harvard.edu/abs/2014A&A...571A..16P} {571, A16}

\bibitem[\protect\citeauthoryear{Polisensky \& Ricotti}{Polisensky \&
  Ricotti}{2011}]{Polisensky2011}
Polisensky E.,  Ricotti M.,  2011, \mn@doi [Phys. Rev. D]
  {10.1103/PhysRevD.83.043506}, 83, 043506

\bibitem[\protect\citeauthoryear{{Primack}}{{Primack}}{2012}]{Primack2012}
{Primack} J.~R.,  2012, \mn@doi [Annalen der Physik] {10.1002/andp.201200077},
  \href {https://ui.adsabs.harvard.edu/abs/2012AnP...524..535P} {524, 535}

\bibitem[\protect\citeauthoryear{{Sales}, {Wetzel}  \& {Fattahi}}{{Sales}
  et~al.}{2022}]{Sales2022}
{Sales} L.~V.,  {Wetzel} A.,   {Fattahi} A.,  2022, \mn@doi [Nature Astronomy]
  {10.1038/s41550-022-01689-w}, \href
  {https://ui.adsabs.harvard.edu/abs/2022NatAs...6..897S} {6, 897}

\bibitem[\protect\citeauthoryear{{Sanders} \& {McGaugh}}{{Sanders} \&
  {McGaugh}}{2002}]{Sanders2002}
{Sanders} R.~H.,  {McGaugh} S.~S.,  2002, \mn@doi [\araa]
  {10.1146/annurev.astro.40.060401.093923}, \href
  {https://ui.adsabs.harvard.edu/abs/2002ARA&A..40..263S} {40, 263}

\bibitem[\protect\citeauthoryear{{Schaller} et~al.,}{{Schaller}
  et~al.}{2015}]{Schaller2015}
{Schaller} M.,  et~al., 2015, \mn@doi [\mnras] {10.1093/mnras/stv1067}, \href
  {https://ui.adsabs.harvard.edu/abs/2015MNRAS.451.1247S} {451, 1247}

\bibitem[\protect\citeauthoryear{{Schaye} et~al.,}{{Schaye}
  et~al.}{2015}]{Schaye2015}
{Schaye} J.,  et~al., 2015, \mn@doi [\mnras] {10.1093/mnras/stu2058}, \href
  {https://ui.adsabs.harvard.edu/abs/2015MNRAS.446..521S} {446, 521}

\bibitem[\protect\citeauthoryear{{Schaye} et~al.,}{{Schaye}
  et~al.}{2023}]{Schaye2023}
{Schaye} J.,  et~al., 2023, \mn@doi [\mnras] {10.1093/mnras/stad2419}, \href
  {https://ui.adsabs.harvard.edu/abs/2023MNRAS.tmp.2384S} {p. stad2419}

\bibitem[\protect\citeauthoryear{{Schmidt}, {Engels}, {Niemeyer}  \&
  {Almgren}}{{Schmidt} et~al.}{2016}]{Schmidt2016}
{Schmidt} W.,  {Engels} J.~F.,  {Niemeyer} J.~C.,   {Almgren} A.~S.,  2016,
  \mn@doi [\mnras] {10.1093/mnras/stw632}, \href
  {https://ui.adsabs.harvard.edu/abs/2016MNRAS.459..701S} {459, 701}

\bibitem[\protect\citeauthoryear{{Schmidt}, {Byrohl}, {Engels}, {Behrens}  \&
  {Niemeyer}}{{Schmidt} et~al.}{2017}]{Schmidt2017}
{Schmidt} W.,  {Byrohl} C.,  {Engels} J.~F.,  {Behrens} C.,   {Niemeyer} J.~C.,
   2017, \mn@doi [\mnras] {10.1093/mnras/stx1274}, \href
  {https://ui.adsabs.harvard.edu/abs/2017MNRAS.470..142S} {470, 142}

\bibitem[\protect\citeauthoryear{{Sefusatti}, {Crocce}, {Scoccimarro}  \&
  {Couchman}}{{Sefusatti} et~al.}{2016}]{Sefusatti2016}
{Sefusatti} E.,  {Crocce} M.,  {Scoccimarro} R.,   {Couchman} H.~M.~P.,  2016,
  \mn@doi [\mnras] {10.1093/mnras/stw1229}, \href
  {https://ui.adsabs.harvard.edu/abs/2016MNRAS.460.3624S} {460, 3624}

\bibitem[\protect\citeauthoryear{{Simpson}, {Grand}, {G{\'o}mez}, {Marinacci},
  {Pakmor}, {Springel}, {Campbell}  \& {Frenk}}{{Simpson}
  et~al.}{2018}]{Simpson2018}
{Simpson} C.~M.,  {Grand} R. J.~J.,  {G{\'o}mez} F.~A.,  {Marinacci} F.,
  {Pakmor} R.,  {Springel} V.,  {Campbell} D. J.~R.,   {Frenk} C.~S.,  2018,
  \mn@doi [\mnras] {10.1093/mnras/sty774}, \href
  {https://ui.adsabs.harvard.edu/abs/2018MNRAS.478..548S} {478, 548}

\bibitem[\protect\citeauthoryear{{Sorini}, {Dav{\'e}}, {Cui}  \&
  {Appleby}}{{Sorini} et~al.}{2022}]{Sorini2022}
{Sorini} D.,  {Dav{\'e}} R.,  {Cui} W.,   {Appleby} S.,  2022, \mn@doi [\mnras]
  {10.1093/mnras/stac2214}, \href
  {https://ui.adsabs.harvard.edu/abs/2022MNRAS.516..883S} {516, 883}

\bibitem[\protect\citeauthoryear{Springel}{Springel}{2005}]{Springel2005}
Springel V.,  2005, \mn@doi [\mnras] {10.1111/j.1365-2966.2005.09655.x}, 364,
  1105

\bibitem[\protect\citeauthoryear{Springel}{Springel}{2010}]{Springel2010}
Springel V.,  2010, \mn@doi [\mnras] {10.1111/j.1365-2966.2009.15715.x}, 401,
  791

\bibitem[\protect\citeauthoryear{{Springel} et~al.,}{{Springel}
  et~al.}{2018}]{Springel2018}
{Springel} V.,  et~al., 2018, \mn@doi [\mnras] {10.1093/mnras/stx3304}, \href
  {https://ui.adsabs.harvard.edu/abs/2018MNRAS.475..676S} {475, 676}

\bibitem[\protect\citeauthoryear{{Sunseri}, {Li}  \& {Liu}}{{Sunseri}
  et~al.}{2023}]{Sunseri2023}
{Sunseri} J.,  {Li} Z.,   {Liu} J.,  2023, \mn@doi [\prd]
  {10.1103/PhysRevD.107.023514}, \href
  {https://ui.adsabs.harvard.edu/abs/2023PhRvD.107b3514S} {107, 023514}

\bibitem[\protect\citeauthoryear{{Takahashi}, {Nishimichi}, {Namikawa},
  {Taruya}, {Kayo}, {Osato}, {Kobayashi}  \& {Shirasaki}}{{Takahashi}
  et~al.}{2020}]{Takahashi2020}
{Takahashi} R.,  {Nishimichi} T.,  {Namikawa} T.,  {Taruya} A.,  {Kayo} I.,
  {Osato} K.,  {Kobayashi} Y.,   {Shirasaki} M.,  2020, \mn@doi [\apj]
  {10.3847/1538-4357/ab908d}, \href
  {https://ui.adsabs.harvard.edu/abs/2020ApJ...895..113T} {895, 113}

\bibitem[\protect\citeauthoryear{{The EAGLE team}}{{The EAGLE
  team}}{2017}]{EAGLE2017}
{The EAGLE team} 2017, \mn@doi [arXiv e-prints] {10.48550/arXiv.1706.09899},
  \href {https://ui.adsabs.harvard.edu/abs/2017arXiv170609899T} {p.
  arXiv:1706.09899}

\bibitem[\protect\citeauthoryear{Torrence \& Compo}{Torrence \&
  Compo}{1998}]{Torrence1998}
Torrence C.,  Compo G.~P.,  1998, \mn@doi [Bull. Amer. Meteor.]
  {10.1175/1520-0477(1998)079<0061:APGTWA>2.0.CO;2}, 79, 61

\bibitem[\protect\citeauthoryear{Torrey, Vogelsberger, Genel, Sijacki, Springel
   \& Hernquist}{Torrey et~al.}{2014}]{Torrey2014}
Torrey P.,  Vogelsberger M.,  Genel S.,  Sijacki D.,  Springel V.,   Hernquist
  L.,  2014, \mn@doi [\mnras] {10.1093/mnras/stt2295}, 438, 1985

\bibitem[\protect\citeauthoryear{{Villaescusa-Navarro}
  et~al.,}{{Villaescusa-Navarro} et~al.}{2021}]{VillaescusaNavarro2021}
{Villaescusa-Navarro} F.,  et~al., 2021, \mn@doi [\apj]
  {10.3847/1538-4357/abf7ba}, \href
  {https://ui.adsabs.harvard.edu/abs/2021ApJ...915...71V} {915, 71}

\bibitem[\protect\citeauthoryear{Vogelsberger, Genel, Sijacki, Torrey, Springel
   \& Hernquist}{Vogelsberger et~al.}{2013}]{Vogelsberger2013}
Vogelsberger M.,  Genel S.,  Sijacki D.,  Torrey P.,  Springel V.,   Hernquist
  L.,  2013, \mn@doi [\mnras] {10.1093/mnras/stt1789}, 436, 3031

\bibitem[\protect\citeauthoryear{Vogelsberger et~al.,}{Vogelsberger
  et~al.}{2014}]{Vogelsberger2014}
Vogelsberger M.,  et~al., 2014, \mn@doi [\nat] {10.1038/nature13316}, 509, 177

\bibitem[\protect\citeauthoryear{{Vogelsberger}, {Marinacci}, {Torrey}  \&
  {Puchwein}}{{Vogelsberger} et~al.}{2020}]{Vogelsberger2020}
{Vogelsberger} M.,  {Marinacci} F.,  {Torrey} P.,   {Puchwein} E.,  2020,
  \mn@doi [Nature Reviews Physics] {10.1038/s42254-019-0127-2}, \href
  {https://ui.adsabs.harvard.edu/abs/2020NatRP...2...42V} {2, 42}

\bibitem[\protect\citeauthoryear{{Voit}}{{Voit}}{2005}]{Voit2005}
{Voit} G.~M.,  2005, \mn@doi [Reviews of Modern Physics]
  {10.1103/RevModPhys.77.207}, \href
  {https://ui.adsabs.harvard.edu/abs/2005RvMP...77..207V} {77, 207}

\bibitem[\protect\citeauthoryear{Wang \& He}{Wang \& He}{2021}]{Wang2021}
Wang Y.,  He P.,  2021, \mn@doi [Communications in Theoretical Physics]
  {10.1088/1572-9494/ac10be}, \href
  {https://ui.adsabs.harvard.edu/abs/2021CoTPh..73i5402W} {73, 095402}

\bibitem[\protect\citeauthoryear{Wang \& He}{Wang \& He}{2022}]{Wang2022b}
Wang Y.,  He P.,  2022, \mn@doi [\apj] {10.3847/1538-4357/ac7a3d}, 934, 112

\bibitem[\protect\citeauthoryear{Wang \& He}{Wang \& He}{2023}]{Wang2023}
Wang Y.,  He P.,  2023, \mn@doi [RAS Techniques and Instruments]
  {10.1093/rasti/rzad020}, 2, 307

\bibitem[\protect\citeauthoryear{{Wang}, {Yang}  \& {He}}{{Wang}
  et~al.}{2022}]{Wang2022a}
{Wang} Y.,  {Yang} H.-Y.,   {He} P.,  2022, \mn@doi [\apj]
  {10.3847/1538-4357/ac752c}, \href
  {https://ui.adsabs.harvard.edu/abs/2022ApJ...934...77W} {934, 77}

\bibitem[\protect\citeauthoryear{{Weinberg}, {Bullock}, {Governato}, {Kuzio de
  Naray}  \& {Peter}}{{Weinberg} et~al.}{2015}]{Weinberg2015}
{Weinberg} D.~H.,  {Bullock} J.~S.,  {Governato} F.,  {Kuzio de Naray} R.,
  {Peter} A.~H.,  2015, \mn@doi [Proceedings of the National Academy of
  Sciences] {10.1073/pnas.1308716112}, 112, 12249

\bibitem[\protect\citeauthoryear{Weinberger et~al.,}{Weinberger
  et~al.}{2018}]{Weinberger2018}
Weinberger R.,  et~al., 2018, \mn@doi [\mnras] {10.1093/mnras/sty1733}, 479,
  4056

\bibitem[\protect\citeauthoryear{{Yang}, {He}, {Zhu}  \& {Feng}}{{Yang}
  et~al.}{2020}]{Yang2020}
{Yang} H.-Y.,  {He} P.,  {Zhu} W.,   {Feng} L.-L.,  2020, \mn@doi [\mnras]
  {10.1093/mnras/staa2666}, \href
  {https://ui.adsabs.harvard.edu/abs/2020MNRAS.498.4411Y} {498, 4411}

\bibitem[\protect\citeauthoryear{{Yang}, {Wang}, {He}, {Zhu}  \& {Feng}}{{Yang}
  et~al.}{2022}]{Yang2022}
{Yang} H.-Y.,  {Wang} Y.,  {He} P.,  {Zhu} W.,   {Feng} L.-L.,  2022, \mn@doi
  [\mnras] {10.1093/mnras/stab3062}, \href
  {https://ui.adsabs.harvard.edu/abs/2022MNRAS.509.1036Y} {509, 1036}

\bibitem[\protect\citeauthoryear{{Zhu}, {Feng}  \& {Fang}}{{Zhu}
  et~al.}{2010}]{Zhu2010}
{Zhu} W.,  {Feng} L.-l.,   {Fang} L.-Z.,  2010, \mn@doi [\apj]
  {10.1088/0004-637X/712/1/1}, \href
  {https://ui.adsabs.harvard.edu/abs/2010ApJ...712....1Z} {712, 1}

\bibitem[\protect\citeauthoryear{{Zhuravleva} et~al.,}{{Zhuravleva}
  et~al.}{2014}]{Zhuravleva2014}
{Zhuravleva} I.,  et~al., 2014, \mn@doi [\nat] {10.1038/nature13830}, \href
  {https://ui.adsabs.harvard.edu/abs/2014Natur.515...85Z} {515, 85}

\bibitem[\protect\citeauthoryear{{de Blok}}{{de Blok}}{2010}]{Blok2010}
{de Blok} W.~J.~G.,  2010, \mn@doi [Advances in Astronomy]
  {10.1155/2010/789293}, \href
  {https://ui.adsabs.harvard.edu/abs/2010AdAst2010E...5D} {2010, 789293}

\bibitem[\protect\citeauthoryear{{van Daalen}, {Schaye}, {Booth}  \& {Dalla
  Vecchia}}{{van Daalen} et~al.}{2011}]{vanDaalen2011}
{van Daalen} M.~P.,  {Schaye} J.,  {Booth} C.~M.,   {Dalla Vecchia} C.,  2011,
  \mn@doi [\mnras] {10.1111/j.1365-2966.2011.18981.x}, \href
  {https://ui.adsabs.harvard.edu/abs/2011MNRAS.415.3649V} {415, 3649}

\bibitem[\protect\citeauthoryear{{van Daalen}, {McCarthy}  \& {Schaye}}{{van
  Daalen} et~al.}{2020}]{vanDaalen2020}
{van Daalen} M.~P.,  {McCarthy} I.~G.,   {Schaye} J.,  2020, \mn@doi [\mnras]
  {10.1093/mnras/stz3199}, \href
  {https://ui.adsabs.harvard.edu/abs/2020MNRAS.491.2424V} {491, 2424}

\makeatother
\end{thebibliography}
\bibliographystyle{mnras}

\appendix

\section{Tests of different mass assignment schemes}
\label{sec:assign_schemes}

\begin{figure}
  \centerline{\includegraphics[width=0.45\textwidth]{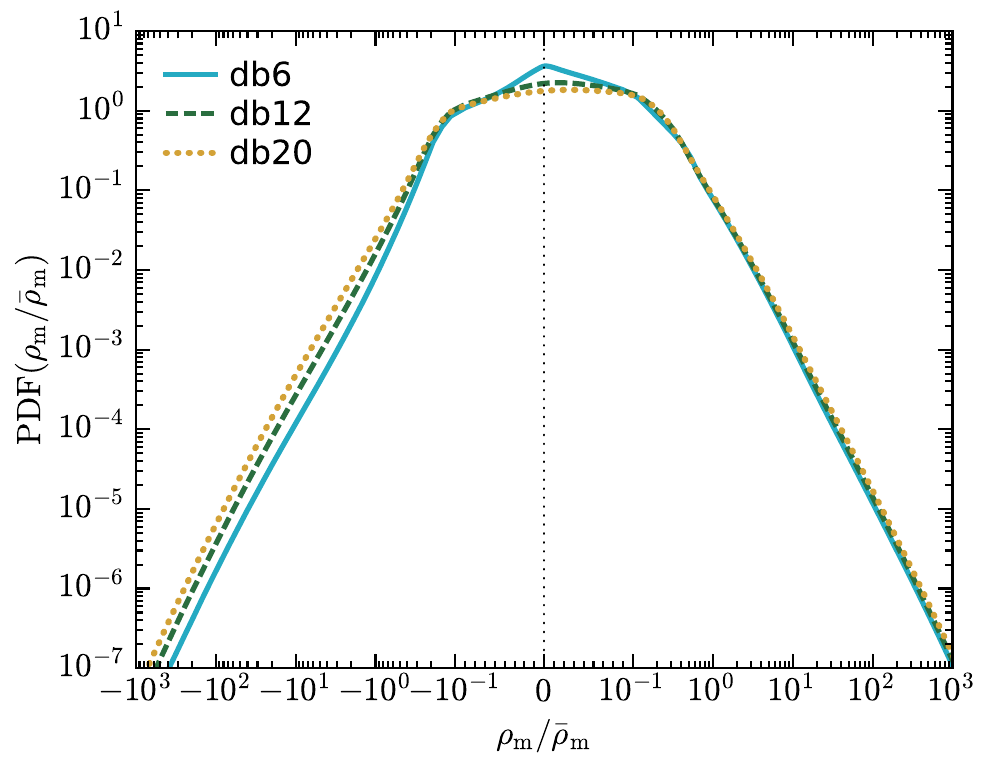}}
  \caption{ The probability distribution function of the density field, $\mathrm{PDF}(\rho_\mathrm{m}/\bar\rho_\mathrm{m})$, measured from the TNG100 simulation by using the Daubechies db6 (solid), db12 (dashed) and db20 (dotted) scale functions, respectively.}
  \label{fig:prob_dis_func}
\end{figure}

To minimize the smearing and aliasing effects on the FPS or other Fourier-based statistics, \citet{Cui2008} found that the scale functions of Daubechies wavelet transformations are the optimal mass assignment window functions. However, does this scheme yield a physical density field in real space? The answer is no, as shown in Fig. \ref{fig:prob_dis_func}, in which we examine the probability distribution functions of the total matter density fields obtained by using three types of Daubechies scale functions \citep[db6, db12, and db20, see][]{Daubechies1992, Hand2018}. It can be seen that Daubechies windows lead to too many unphysical values of $\rho_\mathrm{m}/\bar\rho_\mathrm{m}<0$.

\begin{figure}
  \centerline{\includegraphics[width=0.49\textwidth]{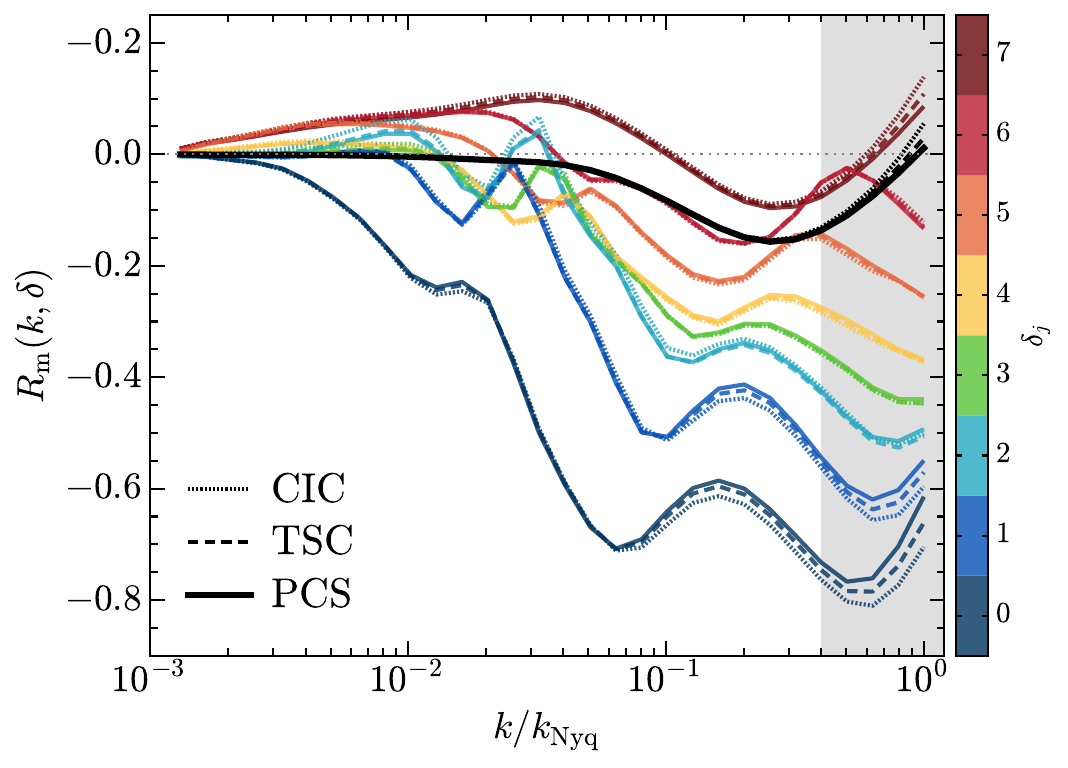}}
  \caption{The relative difference of the total matter env-WPS between the hydrodynamic and DMO runs of TNG100, computed with the CIC (dotted), TSC (dashed), and PCS (solid) schemes. For comparison, the black lines denote the relative difference of the global-WPS. The gray shaded region denotes the scale range of $k>0.4k_\mathrm{Nyq}$.}
  \label{fig:envwps_TNG100_windows}
\end{figure}

As is well known, we can get the physical density field with $\rho_\mathrm{m}/\bar\rho_\mathrm{m}\geqslant0$ by using traditional mass assignment window functions, e.g. the CIC (2nd-order), TSC (3rd-order), and PCS (4th-order) \citep{Sefusatti2016}. Then we test different choices of window functions and investigate how the relative differences $R_\mathrm{m}(k)$ and $R_\mathrm{m}(k,\delta)$ vary between them, the results of which are presented in Fig. \ref{fig:envwps_TNG100_windows}. 
Most of the variation is seen passing from CIC to TSC, whereas the results converge for higher order schemes. We also see that the effects of windows on $R_\mathrm{m}(k)$ are marginal, while the effects on $R_\mathrm{m}(k,\delta)$ are a bit more noticeable in the extremely underdense environments on small scales. Consequently, we use the PCS scheme to compute the density field, and consider only the measurements on scales of $k\leqslant0.4k_\mathrm{Nyq}$ to avoid numerical effects, e.g. smearing, aliasing, and shot noise. 

\section{An example: the total matter env-WPS of TNG100}
\label{sec:example_envwps}

\begin{figure}
	\centerline{\includegraphics[width=0.48\textwidth]{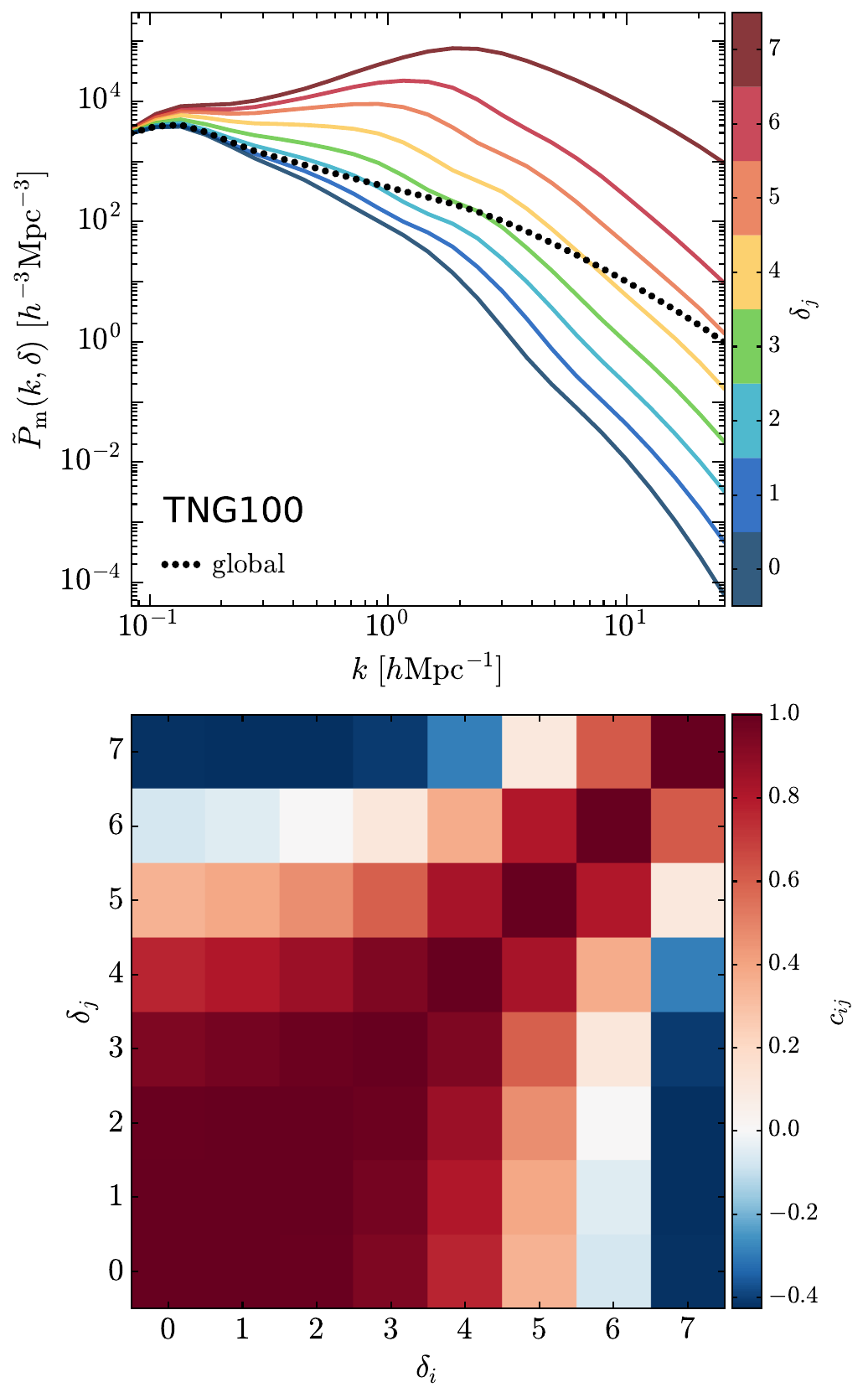}}
	\caption{The env-WPS and its correlation coefficients. \textit{Top panel}: the total matter env-WPS of the TNG100 hydrodynamic run. The global-WPS is also plotted and is denoted by the black dotted line. \textit{Bottom panel}: Correlation coefficients of the env-WPS, defined by $c_{ij}\equiv C_{ij}/\sqrt{C_{ii}C_{jj}}$, in which $C_{ij}$ is the $ij$-th element of the covariance matrix, computed by $C_{ij}=\frac{1}{N_k-1}\sum_n [\tilde{P}_\mathrm{m}(k_n,\delta_i)-\tilde{P}_\mathrm{m}(\delta_i)][\tilde{P}_\mathrm{m}(k_n,\delta_j)-\tilde{P}_\mathrm{m}(\delta_j)]$, where $\tilde{P}_\mathrm{m}(\delta_i)$ is the mean of $\tilde{P}_\mathrm{m}(k_n,\delta_i)$ over scales.}
	\label{fig:envwps_TNG100_hydro_z0}
\end{figure}

We present here an example that we hope will help the reader to gain some intuition about the env-WPS. As shown in the top panel of Fig. \ref{fig:envwps_TNG100_hydro_z0}, the env-WPS increases monotonically with increasing density at almost all scales, while gradually converges to the global-WPS at large scales. This result manifests the non-Gaussian feature developed by the nonlinear gravitational clustering leading to the growth of overdensities. We also estimate the correlation coefficients of the env-WPSs between density environments, which are shown in the bottom panel of Fig. \ref{fig:envwps_TNG100_hydro_z0}. It can be seen that the underdense environments (i.e. $\delta_0$, $\delta_1$, and $\delta_2$ ) are fully correlated with each other, suggesting that the underdensities ($\delta<0$) do not need to be further segmented. The correlation between different environments decreases with increasing density. Especially, the densest environment $\delta_7$ is clearly anti-correlated with environments of $\delta_{0\leq j\leq4}$.

\section{Baryonic effects as a function of the local dark matter density}
\label{sec:baryonic_eff_dm}

\begin{figure}
	\centerline{\includegraphics[width=0.48\textwidth]{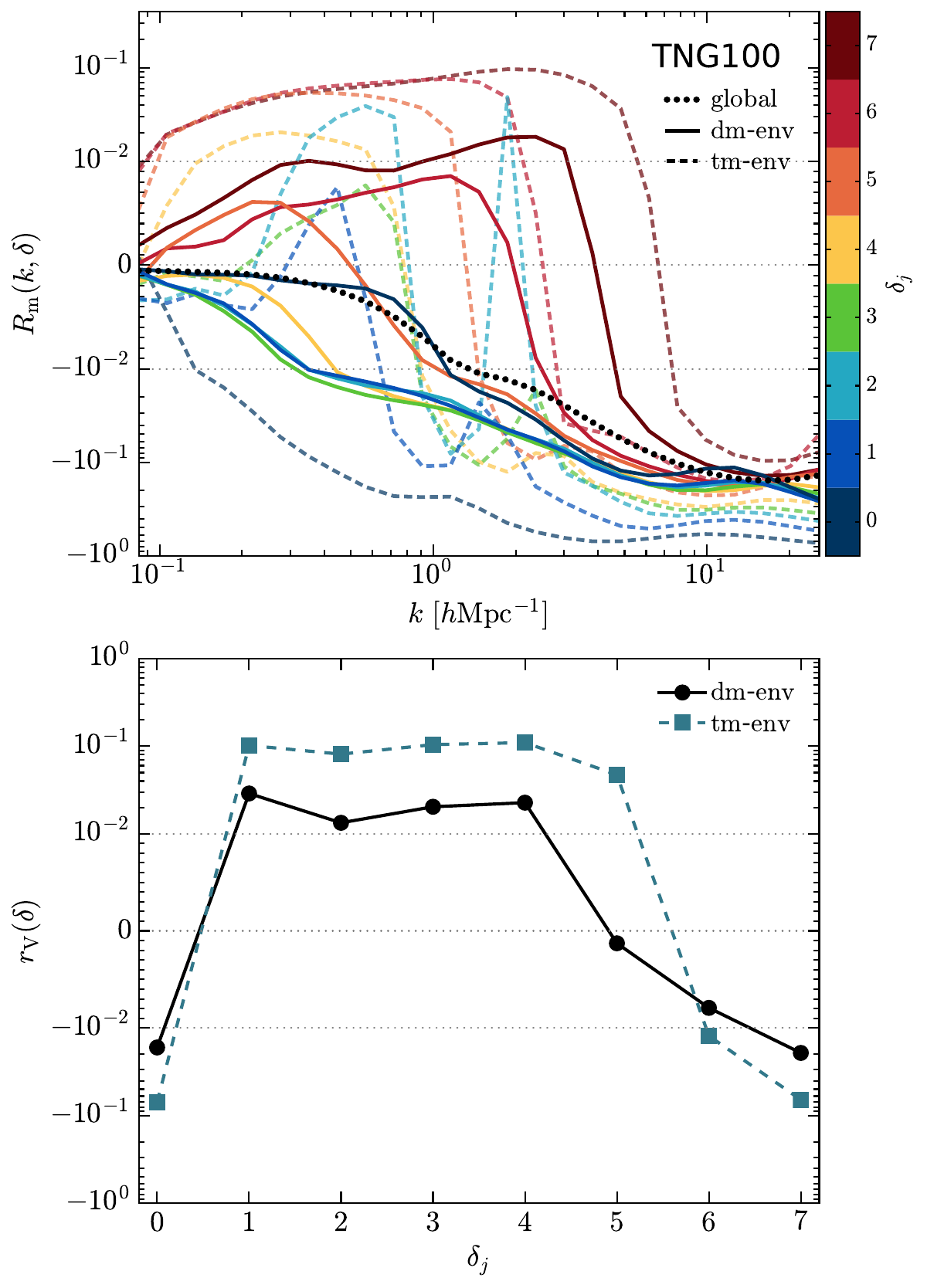}}
	\caption{Baryonic effects as a function of scale and local dark matter density. \textit{Top panel}: The relative difference of the total matter env-WPS between the hydrodynamic and DMO runs of TNG100, computed based on the total matter (tm-env, dashed lines) and dark matter (dm-env, solid lines) density environments, respectively. The relative difference of the total matter global-WPS is plotted as the black dotted line. \textit{Bottom panel}: the relative difference of the environment volume fraction between the hydrodynamic and DMO runs. }
	\label{fig:Renvwps_TNG100_dmenv}
\end{figure}

Throughout our analysis, we have explored baryonic effects by dividing the space into different environments based on the total matter density field (referred to as tm-env), which can be instructive for cosmic shear observations that directly probe the total mass between us and background galaxies. Nevertheless, it might be interesting to repeat the calculation for the hydrodynamic runs as a function of the local dark matter density (referred to as dm-env), the results of which are shown in Fig. \ref{fig:Renvwps_TNG100_dmenv}. We find that both the enhancement of extreme overdense environments and the suppression of underdense environments are weakend. Namely, the baryonic effects on the env-WPS are closer to the global one than in the case of tm-env. This can be explained by the fact that the distribution of dark matter is less affected by baryons, as can be seen from the fact that the volume fraction of the environment changes by less than a few percent.

\section{The residual cosmic variance}
\label{sec:cv}

\begin{figure}
	\centerline{\includegraphics[width=0.49\textwidth]{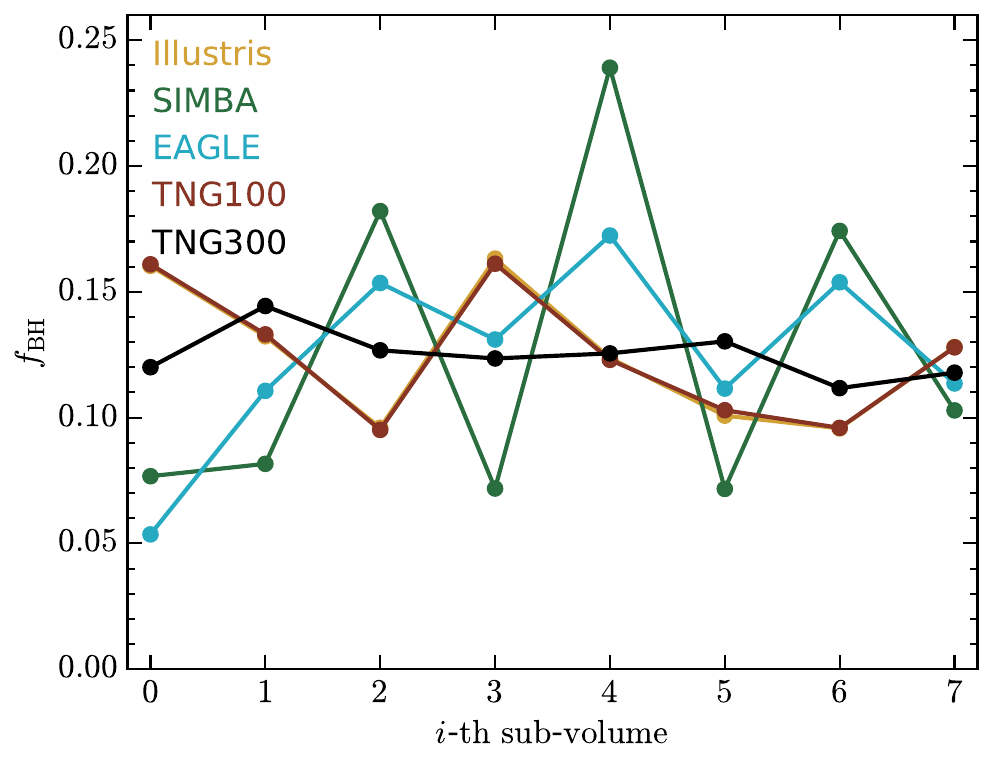}}
	\caption{The number fraction of black hole particles contained in the sub-volume, $f_\mathrm{BH}=N_{\mathrm{BH},\mathrm{sub}}/N_{\mathrm{BH}}$, where $N_{\mathrm{BH},\mathrm{sub}}$ and $N_{\mathrm{BH}}$ are the number of black hole particles in the sub-volume and in the full volume, respectively.}
	\label{fig:fraction_black_holes}
\end{figure}

Here we describe in more detail how the residual cosmic variance effects are estimated. First, we perform the CWT of the density field at $N_k$ scales. Then we divide the CWT field into $N_\mathrm{sub}=8$ equal-sized subfields at each scale, and calculate the relative differences within each subfield, i.e. $R_{\text{m,}i}\left( k,\delta \right) =\tilde{P}_{\text{m,}i}\left( k,\delta \right)/\tilde{P}_{\text{DMO,}i}\left( k,\delta \right)-1$ and $R_{\text{m,}i}\left( k \right) =\tilde{P}_{\text{m,}i}\left( k \right)/\tilde{P}_{\text{DMO,}i}\left( k \right)-1$, where ``$i$" denotes the $i$-th subfield. Finally, if we use $R_{\text{m,med}}$ to denote the median of relative differences between the different subfields, then the cosmic variances are indicated by the 1-$\sigma$ (68 percentile) dispersion around the median, $R_{\text{m,med}}$, as shown in Figs. \ref{fig:tot_globalWPS} and \ref{fig:tot_envWPS}. We find that all relative differences measured from the total simulation volume are within the 1-$\sigma$ uncertainty, except for those at $k\lesssim 1\ h\mathrm{Mpc}^{-1}$ and in $\delta_{j\geqslant6}$ for SIMBA. Given this scale and environment, this result may reflect the inhomogeneous distribution of AGNs in SIMBA. To check this, we count the number fraction of black hole particles in each sub-volume, the results of which are shown in Fig.~\ref{fig:fraction_black_holes}. Compared to the other simulations, it can be seen that the dispersion of the number fractions across sub-volumes is much larger in SIMBA, demonstrating the inhomogeneity of the AGNs.

\end{document}